%% file: main.tex
\documentclass[sn-mathphys-num]{sn-jnl}

\usepackage{graphicx}%
\usepackage{multirow}%
\usepackage{amsmath,amssymb,amsfonts}%
\usepackage{amsthm}%
\usepackage{mathrsfs}%
\usepackage[title]{appendix}%
\usepackage{xcolor}%
\usepackage{textcomp}%
\usepackage{manyfoot}%
\usepackage{booktabs}%
\usepackage{algorithm}%
\usepackage{algorithmicx}%
\usepackage{algpseudocode}%
\usepackage{listings}%
\usepackage{glossaries}
\usepackage{cleveref}
\usepackage{braket}
\usepackage{caption}
\usepackage{subcaption}
\usepackage{url}
\usepackage{placeins}

\newcommand{\samplemean}{\bar{x}}
\newcommand{\samplesd}{s}
\newcommand{\expertexpert}{\textsc{Expert-Expert}}
\newcommand{\expertllm}{\textsc{Expert-\acrshort{llm}}}
\newcommand{\llmllm}{\textsc{\acrshort{llm}-\acrshort{llm}}}
\newcommand{\llmexpert}{\textsc{\acrshort{llm}-Expert}}

\input{glossary}

\begin{document}

\title[Only a Tip]{LLM-Generated Tips Rival Expert-Created Tips in Helping Students Answer Quantum-Computing Questions}


\author*[1,2]{\fnm{Lars} \sur{Krupp}}\email{lars.krupp@dfki.de}

\author[3]{\fnm{Jonas} \sur{Bley}}\email{jonas.bley@rptu.de}

\author[4]{\fnm{Isacco} \sur{Gobbi}}\email{isacco.gobbi@itwm.fraunhofer.de}

\author[4]{\fnm{Alexander} \sur{Geng}}\email{alexander.geng@itwm.fraunhofer.de}

\author[4]{\fnm{Sabine} \sur{Müller}}\email{sabine.mueller@itwm.fraunhofer.de}

\author[1]{\fnm{Sungho} \sur{Suh}}\email{sungho.suh@dfki.de}

\author[4]{\fnm{Ali} \sur{Moghiseh}}\email{ali.moghiseh@itwm.fraunhofer.de}

\author[4]{\fnm{Arcesio} \sur{Castaneda Medina}}\email{arcesio.castaneda.medina@itwm.fraunhofer.de}

\author[4,5]{\fnm{Valeria} \sur{Bartsch}}\email{valeria.bartsch@cml.fraunhofer.de}

\author[3]{\fnm{Artur} \sur{Widera}}\email{widera@rhrk.uni-kl.de}

\author[3]{\fnm{Herwig} \sur{Ott}}\email{ott@physik.uni-kl.de}

\author[1,2]{\fnm{Paul} \sur{Lukowicz}}\email{paul.lukowicz@dfki.de}

\author[1,2]{\fnm{Jakob} \sur{Karolus}}\email{jakob.karolus@dfki.de}

\author[1,2,3]{\fnm{Maximilian} \sur{Kiefer-Emmanouilidis}}\email{maxkiefer@physik.uni-kl.de}

\affil[1]{\orgdiv{Embedded Intelligence}, \orgname{German Research Center for Artificial Intelligence}, \orgaddress{\city{Kaiserslautern}, \postcode{67663}, \country{Germany}}}

\affil[2]{\orgdiv{Department of Computer Science and Research Initiative QC-AI}, \orgname{RPTU Kaiserslautern-Landau}, \orgaddress{\city{Kaiserslautern}, \postcode{67663}, \country{Germany}}}

\affil[3]{\orgdiv{Department of Physics and Research Center OPTIMAS}, \orgname{RPTU Kaiserslautern-Landau}, \orgaddress{\city{Kaiserslautern}, \postcode{67663}, \country{Germany}}}

\affil[4]{\orgname{Fraunhofer Institute for Industrial Mathematics},
\orgaddress{\city{Kaiserslautern}, \postcode{67663}, \country{Germany}}}

\affil[5]{\orgname{Fraunhofer Center for Maritime Logistics and Services},
\orgaddress{\city{Hamburg}, \postcode{21073}, \country{Germany}}}

\abstract{
Individual teaching is among the most successful ways to impart knowledge. Yet, sometimes, this method is not feasible due to large numbers of students per educator, especially in specialized subjects. The subject of quantum computing serves as a prime example facing this issue, due to its growing media presence and anticipation of future significance.
Alleviating high workloads for teachers, often accompanied with individual teaching, is crucial for continuous high quality education. In this regard, leveraging Large Language Models (\acrshort{llm}s) such as GPT-4 to generate educational content can be valuable support.
To evaluate this concept, we conducted two complementary studies exploring the feasibility of using GPT-4 to automatically generate tips for students. Our main study consisted of a between-subject survey in which students (N = 46) solved four multiple-choice quantum computing questions with either the help of expert-created or \acrshort{llm}-generated tips. To correct for possible biases towards \acrshort{llm}s, we introduced two additional deception conditions, making some participants believe that they were given expert-created tips, when they were in fact given \acrshort{llm}-generated tips and vice versa. Our second study (N = 23) aimed to directly compare the \acrshort{llm}-generated and expert-created tips, evaluating their quality, correctness and helpfulness. Both experienced educators and students took part in the tip evaluation study.
The results from both studies highlight that \acrshort{llm}-generated tips can be equally useful as expert-created tips. Furthermore, participants in our second study found that the \acrshort{llm}-generated tips were significantly more helpful and pointed better towards relevant concepts than the expert-created tips. However, \acrshort{llm}-generated tips were more prone to be giving away the answer easily. Lastly, we found that participants in the first study performed significantly better in answering the quantum computing questions when given tips labeled as \acrshort{llm}-generated, even if they were created by an expert. We argue that this phenomenon could be a placebo effect of artificial intelligence induced by the participants' biases for \acrshort{llm}-generated content. Ultimately, we contribute that \acrshort{llm}-generated tips are good enough to be used instead of expert tips in the context of quantum computing basics. Our work highlights a direction to reduce teachers' workloads and proposes a paradigm shift for a more individualized way of teaching.


}




\maketitle

\newpage

\section{Introduction}\label{introduction}

For a long time, Quantum Computing (\acrshort{qc}) was a niche area within the field of physics that received little to no attention in other Science, Technology, Engineering \& Mathematics (\acrshort{stem}) fields. The recent hype in quantum technologies also known as 'the second quantum revolution'~\cite{MacFarlane2003, Roberson2021, Scheidsteger2022, Roberson_Raman_Leach_Vilkins_2023} has led to significant advances in hardware~\cite{Arute2019, Kim2023, Aboy2022} and formulations of technological standards~\cite{vanDeventer2022, mazurquantum, jenet2020standards4quantum} in the areas of quantum communication, sensing and computation. Today, \acrshort{qc} is one of the fastest-growing interdisciplinary research areas at the intersection of computer science, mathematics, and physics \cite{Sood2024,coccia2024new,Houssein2022, Preskill2018quantumcomputingin}. Thus, it is no surprise that quantum computing has become an interesting topic for industry and academia~\cite{bova2021commercial,bayerstadler2021industry,greinert_2024_10976836,saez2024quantum}, higher education~\cite{CARBERRY20212065, 10.1145/3481312.3481348,goorney2023frameworkcurriculumtransformationquantum,Goorney2024,Meyer2024}, and even high school education~\cite{angara2020quantum,angara2021teaching,walsh2021piloting,Merzel2024}.


However, the amount of experts and educators is still quite limited and mostly intertwined in academia and large companies who have taken the lead in \acrshort{qc} such as Google, IBM and Microsoft~\cite{Arute2019, Kim2023, Aboy2022, Scheidsteger2022, Microsoft2022, Defianceetf2023}. The initial barrier to learn quantum algorithms is high, demanding experienced educators and individual teaching. This poses a burden for educators, adding to their workload and making \acrshort{qc} mostly inaccessible to learners. Consequently, there is a need to make learning and teaching about \acrshort{qc} more accessible. In our work, we want to address these challenges and investigate ways to support educators.

In this work, we explore the use of generative educational content from Large Language Models (\acrshort{llm}s) as a support mechanism for educators, alleviating some of their workload and allowing for more personalized teaching. \acrshort{llm}s are a promising new technology that is already successfully used in education~\cite{kasneci2023chatgpt}. However, how, if and when \acrshort{llm}s should be used in education and what possible consequences might arise from that use for students and educators is a strongly debated topic that is studied intensively at the moment~\cite{ahmad2023gen}.
While some results demand caution~\cite{krupp2024unreflected, gregorcic2023chatgpt, santos2023enhancing}, others report more neutral~\cite{krupp2023challenges, sallam2023chatgpt} or positive results~\cite{gregorcic2024chatgpt}. This shows a need for research on the use of \acrshort{llm}s for quantum computing education.


\acrshort{llm}s first became a topic of public interest with the publication of ChatGPT in November of 2022~\cite{openaichatgpt2022}. Since then, they have been used successfully in a number of different domains~\cite{latif2024physicsassistant, wermelinger2023using, gomez2023confederacy, thirunavukarasu2023large} and a considerable amount of resources has been invested to improve this technology, supported by large corporations~\cite{geminiteam2024gemini,llama32024, microsoftcollab2023,parmar2024nemotron,templeton2024scaling}.  
This led to significant scientific progress being made in a short amount of time. Now, with models like GPT4~\cite{openai2024gpt4} and Llama3~\cite{llama32024} at the forefront of these developments, many tasks that might have been considered difficult to automate a few years ago, can be done to a satisfying degree by computers. This includes writing code~\cite{thirunavukarasu2023large} and creative writing~\cite{gomez2023confederacy}.
In addition, supporting architectures are being developed, such as \acrfull{rag} ~\cite{lewis2020retrieval}, further improving the quality of model outputs by providing them with the relevant context to answer specific questions, while reducing the likelihood of bad outcomes, such as hallucinations~\cite{gao2024retrievalaugmented}. 

This significant recent interest in \acrshort{llm}s has led to the emergence of many new research questions in the field of education. These can be divided into the discovery of possible applications, the evaluation of practicality and effectiveness, and the discussion of challenges and ethicality~\cite{https://doi.org/10.1111/bjet.13370}. As summarized by \citet{https://doi.org/10.1111/bjet.13370}, possible applications of \acrshort{llm}s in education are vast: They include profiling and labeling (classification and structuring of texts and tasks), detection (e.g., of student's and teacher's behaviors and intents), assessment and grading (formative and summative, and also independent of the teacher), teaching support (in- and out-of-classroom), prediction (of student performance, engagement and growth), knowledge representation (graph construction and tracing), feedback (real time or post-hoc and feedback on feedback), content generation (of questions, tasks, code and replies), and recommendation (e.g., of resources and courses).

Scaffolding, the educational practice of providing support and guidance to students when needed~\cite{belland2017instructional}, is one of these applications. 
It can help to achieve the desired learning outcome by reducing the effort needed to complete non-essential tasks, allowing students to focus their attention on the important tasks.
Providing tips dynamically and individually is a widely used scaffolding method in education~\cite{SCHARTEL201277}. However, this can be time-consuming or might even become unfeasible with growing student-to-educator ratio. \acrshort{llm}s can help alleviate this issue by providing individualized tips on demand~\cite{BAJAJ2018834, 10386291}. 

The use of \acrshort{llm}s is seen as natural in the shift to the so-called smart education~\cite{BAJAJ2018834,10386291}, which is the practice of providing personalized learning opportunities on demand, often associated with digitization~\cite{Zhu2016}. With increasing capability of \acrshort{llm}s, the number of possible educational applications increases and challenge and risk assessments change~\cite{10.1145/3531146.3533088,Milano2023}. Due to this significant change of landscape, there have not yet been many studies assessing the effectiveness of state-of-the-art \acrshort{llm}s for education~\cite{https://doi.org/10.1111/bjet.13370}. In the same sense, a few years ago, using \acrshort{llm}s to aid in the context of quantum education would have been unthinkable.

We contribute to this discourse with two interconnected studies. We evaluated how \acrshort{llm}-generated tips compared to expert-generated tips when answering quantum computing questions in two studies. Our main study simulates a practice scenario for students, tasking them to answer questions about the lecture material providing tips as support. This study was conducted as part of the \acrfull{quip} winter school QUIKSTART 2024~\footnote{\url{https://www.quantum-rlp.de/studierende/sommer-und-winterschulen/}}. The second study complements this initial evaluation by validating the tips in a within-subject design focusing on opinions from both experienced educators and students with regard to the employed tips in the main study. By using the same tips in both studies, we ensure that the results can be aggregated to form a bigger picture.

Our results indicated that \acrshort{llm}-generated tips rivaled those of experts, being perceived as significantly more helpful and focusing more on conveying relevant concepts. Yet, certain design limitations need to be considered, such as a possible placebo effect of artificial intelligence~\cite{kosch2023placebo} and generally more leading tips when generating them with an \acrshort{llm}.



In this work, we take a step towards answering some of the underlying questions concerning the quality, helpfulness, and correctness of \acrshort{llm}-generated educational content for essential teaching methods~\cite{belland2017instructional} such as scaffolding, in the domain of quantum computing. Our findings point towards a paradigm shift of \acrshort{llm}-supported smart education, highlighting its feasibility and providing insights into its successful application.


\section{Methods}\label{methods}
In this work, we intend to answer two research questions through two interconnected studies. In our main study, we asked how GPT4-generated tips compare to expert-created tips in the domain of quantum computing. We conducted this study with participants (N = 46) from the \acrfull{quip} winter school QUIKSTART 2024~\footnote{\url{https://www.quantum-rlp.de/studierende/sommer-und-winterschulen/}}. Employing a between-subject design with the factors \emph{tip creator} and \emph{tip label}, we tasked the students with four questions about quantum computing. Students were given one tip for each question. Tip creators (\acrshort{llm}, expert) and how the tips were labeled (\acrshort{llm}, expert) were subject to the condition assigned to each participant, including two consistent conditions (creator equaled label) and two contradictory conditions (creator and label differed). The students were not made aware of the real tip creator until after completing the study.

To validate our findings, we conducted a complementary tip evaluation study (N = 23). Here we specifically asked people with the relevant competencies (both experienced educators and students) to rate how \acrshort{llm}-generated tips compare to expert-created tips.

\paragraph{RQ1: Can \acrshort{llm}-generated tips be used instead of expert-created tips to help students answer quantum-physics questions?} This research question is composed of two parts.

\textbf{RQ1a: Is there a difference in the score, an objective measure, when using LLM-generated tips compared to expert-created tips?}
To answer this question, we evaluated the scores achieved by students in practice scenario of the main study.  

\textbf{RQ1b: Are there subjective differences in the perception of \acrshort{llm}-generated tips and expert-created tips?}
Here we evaluated the subjective measures of our studies. For the main study, we analyzed whether there are differences in perceived correctness, helpfulness, and quality between our conditions. Additionally, we conducted the tip evaluation study, where we asked students and educators to rate the helpfulness and correctness of the tips and if they gave away the answer or pointed to relevant concepts. This was done without giving them any information on the original tip creator.

\paragraph{RQ2: What are adverse effects of \acrshort{llm}-generated tips?} 
Recent work in human computer interaction has shown, that the mere label of \acrshort{ai} can influence the participants' perception, culminating in a placebo effect if participants' are falsely told that there is an \acrshort{ai} system in place~\cite{kosch2023placebo,kloft2024ai}. To evaluate if such a bias or placebo effect exists when participants receive \acrshort{llm}-generated tips, we introduced two deception conditions to our main study, where tip creator and tip label are not the same. This method allowed us to analyze if the tip label influenced any of our metrics.
Additionally, we used the tip evaluation study to analyze if the \acrshort{llm}-generated tips gave away more of the answers compared to the expert-created questions. This could negatively impact the students learning-gains. Even though we eliminated the possibility to give wrong tips due to hallucination by manually checking each \acrshort{llm}-generated tip, we evaluated the perceived tip correctness and quality in the main study and the tip correctness in the tip evaluation study.

\subsection{Multiple-Choice Question Creation}
\label{sec:questions}


We chose topics related to the basics of quantum computing for our questions and made sure that they were covered at QUIKSTART 2024. This way, we established a common ground of knowledge for all study participants taking part in our main study.

We chose two topics and two difficulty levels within each topic. All questions and their answers can be found in the Appendix~\ref{secA1}. Firstly, the Bloch sphere, a typical graphical visualization of single-qubit systems. However, this representation can still cause some difficulties in learners \cite{Hu_2024}. The first, simpler, question asks learners to identify the state $\ket{\psi}=\frac{1}{\sqrt{2}}\ket{0}+\frac{1}{\sqrt{2}}\ket{1}$ on the Bloch sphere. The second, more advanced, question asks the same for the state $\ket{\psi}=\frac{1}{2}\ket{0}+\frac{\sqrt{3}}{2}e^{i\pi/2}\ket{1}$.

For the second topic, we decided on some basics of two-qubit quantum computing: a gate identity involving two Hadamard gates and two CNOT gates (the simpler question) and the exact outcome of a circuit meant for the creation of a Bell state (the more advanced question). The gate identity asks participants to know that the property of these quantum gates is their own inverse. Bell states are the most basic two-qubit maximally entangled states that are essential for any teaching course about entanglement. The creation process of these states is therefore an important knowledge domain. In the question we used, participants were asked to predict the outcome of the $(X\otimes I) CNOT_{12} (H \otimes I)$ operation on the $\ket{00}$ state. 

\subsection{Tip Creation}
\label{sec:tips}
For each of the four multiple-choice questions of the main study, we acquired one expert and one \acrshort{llm} tip.
The expert tips were created by two quantum physics experts with teaching experience in the field who were giving lectures at the winter school. They had access to the questions, answers and the lecture script but not to the \acrshort{llm}-generated tips, to not influence their creative process. A different quantum physics expert confirmed the correctness of the tips and that they do not give away the answer.

The \acrshort{llm} tips were generated by giving the question, answers, and lecture script explaining the quantum computing topics to GPT4 and prompting it to generate a tip for each question in a \acrshort{rag}-like process. In this way, up to five tips per question were generated. In the next step, we randomly selected a tip for each question. Each of those was later validated by a quantum computing expert.

\subsection{Main Study}
\label{sec:main_study}

\begin{figure}[htb]
    \centering
    \includegraphics[width=0.9\textwidth]{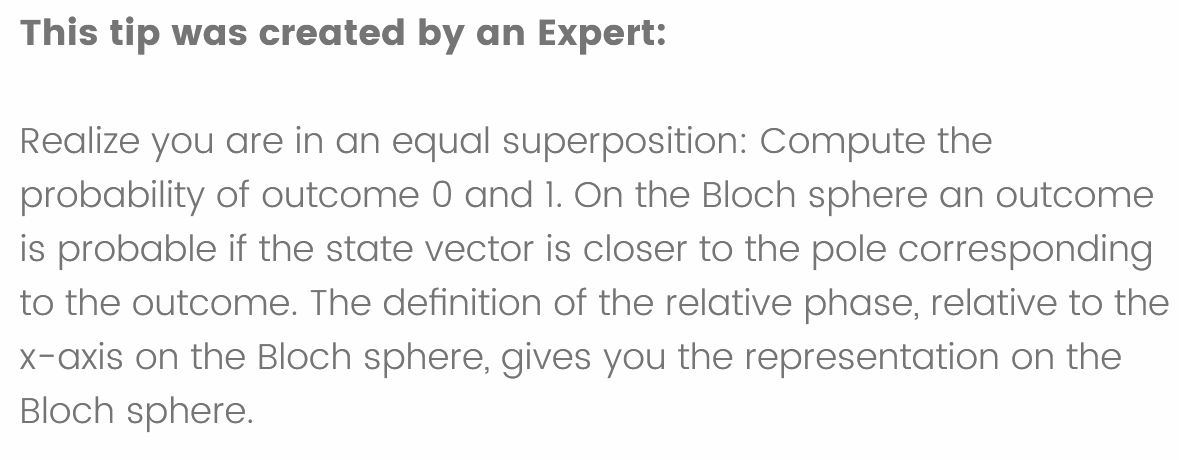}
    \caption{A screenshot of how tips were displayed in the main study.}
    \label{fig:tip_screenshot}
\end{figure}

Our main study was a between-subject survey conducted during QUIKSTART 2024. A winter school conducted specifically for students with a background in machine learning or quantum physics. Participation was entirely voluntary and allowed students that participated in the winter school online or in person and students that did not participate in the winter school to test their quantum computing knowledge in the areas of Bloch spheres and Bell states using multiple-choice questions (\Cref{sec:questions}). Participants were tasked to answer the given questions and were additionally provided with a tip for each of the questions. The type of tip was dependent on the randomly assigned conditions covering two factors: tip creator (\Cref{sec:tips}) and tip label. Hence, a tip could either be created by an expert or the \acrshort{llm}. Likewise, it could either be labeled as being created by an expert or generated by an \acrshort{llm}, yielding a total of four different conditions (see \Cref{fig:tip_screenshot}). We did not include a condition where students received not tip at all. Since our work is based on scaffolding as a teaching method, we already know that providing tips is an effective improvement during learning~\cite{belland2017instructional, SCHARTEL201277}.

Henceforth, we will use the following structure to differentiate between conditions: \textsc{Creator-Label}. In the first condition (\expertexpert) the tip was created by an expert and labeled as such. \llmexpert{} refers to \acrshort{llm}-created tips that were displayed to the participant as if they were created by an expert. \llmllm{} refers to \acrshort{llm}-created tips labeled as such. Finally, \expertllm{} refers to expert-created tips labeled as if they were created by an \acrshort{llm}.

By including the \llmexpert{} and \expertllm{} conditions, and therefore deceiving participants that were assigned to those conditions, we were able to determine the labels' influence on the participants perception of the tip. This allowed us to evaluate the existence of possible placebo effects that have been shown to occur in some settings where humans interact with \acrfull{ai}~\cite{kosch2023placebo}. The deceived participants were informed immediately after answering their demographic questions and shown the correct answers, mitigating possible negative effects of the deception~\cite{boynton2013exploring}. 

We measured the students' scores and further evaluated the perceived usefulness of the displayed tips using custom questions (see \Cref{tab:main_custom_questions}), covering correctness, helpfulness, and quality of the tip. Additionally, we asked participants to rate the perceived difficulty of the question. 
Ethics approval was granted by the Ethics Department of the German Research Center for Artificial Intelligence (\acrshort{dfki}).

\begin{table}[tb]
	\centering
 \caption{helpfulness, quality, and correctness asked for each displayed tip; from \textit{strongly disagree} to \textit{strongly agree}; all visual analog scale (0 to 100).}
	\label{tab:main_custom_questions}
	\begin{tabular}{ll}
		\toprule
		\multicolumn{2}{l}{\textbf{Please rate the following statements regarding the displayed tip.}}\\
		\midrule
		\textbf{CQ1} & The tip was helpful.\\
		\textbf{CQ2} & The quality of the tip was high.\\
		\textbf{CQ3} & The tip was correct.\\
        
		\bottomrule
	\end{tabular}
 
\end{table}



\subsubsection{Main Study Procedure}
\begin{figure}[htb]
    \centering
    \includegraphics[width=\textwidth]{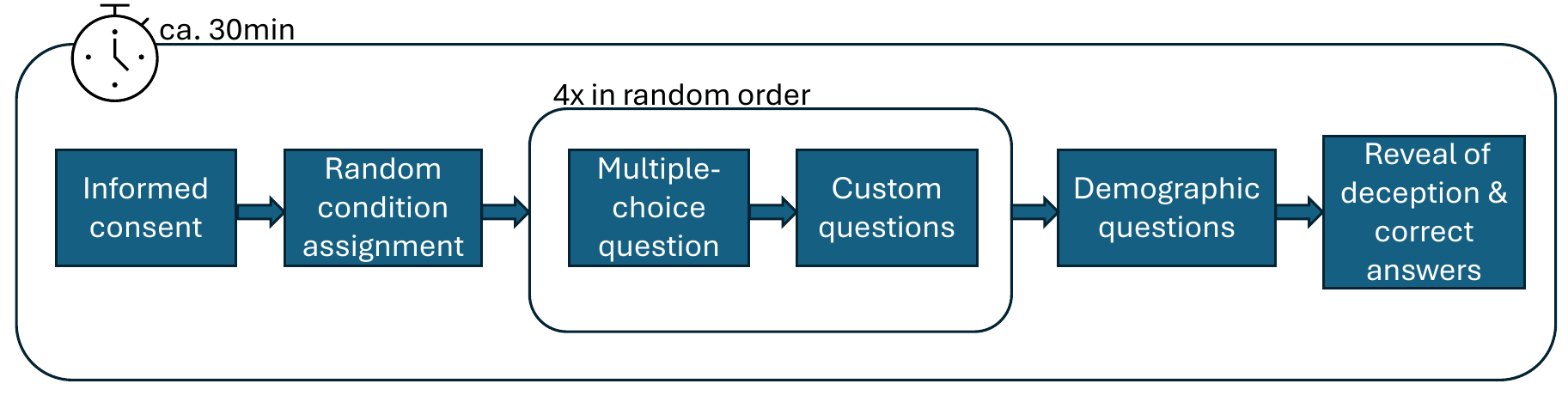}
    \caption{The complete main study procedure, detailing each step.}
    \label{fig:main_study_procedure}
\end{figure}

The complete study procedure is depicted in \Cref{fig:main_study_procedure}. After providing informed consent and being assigned a condition, students were shown the four quantum computing question in randomized order. Each question was displayed individually including its tip. After providing the answer, we inquired about the qualities of the display tips using our custom questions (see \Cref{tab:main_custom_questions}). This process was repeated for each question.

Finally, participants were asked to provide demographics (age, gender), got informed about the deception (if any) and shown the correct answer to each question. The whole study took approximately 30 minutes.

\subsubsection{Main Study Participants}
In total, 46 participants (Age: $\samplemean=28.9\,y$, $\samplesd=7.35\,y$; 35 male, 8 female, 2 no comment, 1 other)\footnote{$\samplemean$ denotes the mean of a sample; $\samplesd$ denotes the standard deviation of a sample} completed the study. Of these participants, 25 attended the winter school in person, 18 online, and 3 did not participate.
Due to the random condition selection, the participants are not evenly distributed onto the four conditions. The participant distribution and demographics can be seen in \Cref{tab:main_study_participant_stats}. 

\begin{table}[]
    \centering
    \caption{The distribution of participants onto the four conditions.}
    \label{tab:main_study_participant_stats}
    \begin{tabular}{ll|l}
    \toprule
        \textbf{Tip creator} &\textbf{Tip label} &  \textbf{\#Participants}\\
        \midrule
        Expert & Expert                 & 9  \\
        \acrshort{llm} & Expert         & 12 \\
        \acrshort{llm} & \acrshort{llm} & 10 \\
        Expert & \acrshort{llm}         & 15 \\
    \bottomrule
    \end{tabular}
    
\end{table}

\subsection{Tip Evaluation Study}
After we completed the main study, we conducted a separate tip evaluation study. The purpose of this study was to form a better understanding of the quality of the tips used in the main study. In this within-subjects study, we had participants rate each tip on a custom scale adjusted to facilitate the evaluation of measures not tested in the main survey (see~\Cref{tab:tip_eval_custom_questions}) and allowed for direct comparison of \acrshort{llm}-generated and expert-created tips. We acquired participants by asking educators (at least postdoctoral researchers with teaching experience in relevant fields) and students from related fields to participate. This way, we made sure that each participant had the necessary knowledge to rate the tips. The study was not part of the official program at QUIKSTART 2024, unlike our main study. Ethics approval was granted by the Ethics Department of the German Research Center for Artificial Intelligence (\acrshort{dfki}).

\begin{table}[htb]
	\centering
  \caption{Additional questions asked for each displayed tip; from \textit{strongly disagree} to \textit{strongly agree}; all visual analog scale (0 to 100).}
	\label{tab:tip_eval_custom_questions}
	\begin{tabular}{ll}
		\toprule
		\multicolumn{2}{l}{\textbf{Please rate the following statements regarding the displayed tip.}}\\
		\midrule
		\textbf{CQ4} & The tip was correct.\\
		\textbf{CQ5} & The tip gave away the correct answer\\
        \textbf{CQ6} & The tip was helpful.\\
        \textbf{CQ7} & The tip effectively points towards relevant concepts.\\
		\bottomrule
	\end{tabular}

\end{table}

\subsubsection{Tip Evaluation Study Procedure}
After providing their informed consent, participants were asked to quantify their perceived knowledge of quantum physics and to answer some demographic questions. After this, the participants were shown each of the four multiple-choice questions used in the main study, their answer options, the associated correct answer, and both their \acrshort{llm}-generated and their expert-created tips. They were not informed about the creator of each tip, only that they should rate and compare them. 
For each tip, participants answered if it was helpful, correct, gave away the correct answer, and effectively points towards relevant concepts on a scale of 0 to 100, with 0 being the maximum disagreement, and 100 the maximum agreement (see~\Cref{tab:tip_eval_custom_questions}). Furthermore, for each question, the participants decided which tip, if any, was better. Finally, we asked the participants to explain their reasoning for that decision. This allowed us to qualitatively evaluate their responses. The whole study took approximately 10 minutes.

\subsubsection{Tip Evaluation Study Participants}
In total 23 participants completed the survey. The study participants were hand-selected to fulfill the required minimum competencies in \acrshort{qc} to be able to rate the tips. We allowed participants from the main study to participate in this study (eight of 23 participated in both studies). They were divided into two groups, students and educators. The 13 participants designated as educators (Age: $\samplemean=42.9\,y$, $\samplesd=7.7\,y$; 10 male, 3 female) have obtained a Ph.D. and are working and teaching in academia in a multitude of related fields. While not all of them are experts in quantum computing, all have a basic level of understanding that allowed them to judge the tips. This allowed us to get the opinion of multiple (at least four) professors on the topic.
Seven educators are working in the area of physics, with two of them specializing in quantum physics, two are working in computer science, two in artificial intelligence, one in mathematics, and one in applied mathematics and physics. The other ten participants were designated as students (Age: $\samplemean=27.6\,y$, $\samplesd=5.16\,y$; 7 male, 3 female). Of them six studied physics (four physics, two quantum physics), three computer science (one quantum machine learning, one machine learning, one computer science) and one gave no comment.

\section{Results}\label{results}
We divided the results section into two parts, one for each of the conducted studies.
\subsection{Main Study Results}
To evaluate the results of the main study, we conducted two-way analyses of variance (\acrshort{anova})~\cite{girden1992anova} using the factors tip creator and tip label. We constructed models for each of the measures in our main study, including the students' score, their rating of our custom questions (see~\Cref{tab:main_custom_questions}) and their rating of the questions difficulty.
In cases where normality of the data was violated, we first used aligned rank transform (\acrshort{art})~\cite{wobbrock2011aligned} on the data. We found no interaction effects regarding the two factors for any dependent variable and report statistics for the additive models only. Effect sizes are
given using $\eta^2$ (Partial Eta Squared): small ($>.01$), medium ($>.06$), large ($>.14$). All statistics are reported in \Cref{additional_results}.

\begin{figure}[htb]
    \centering
    \includegraphics[height=5cm]{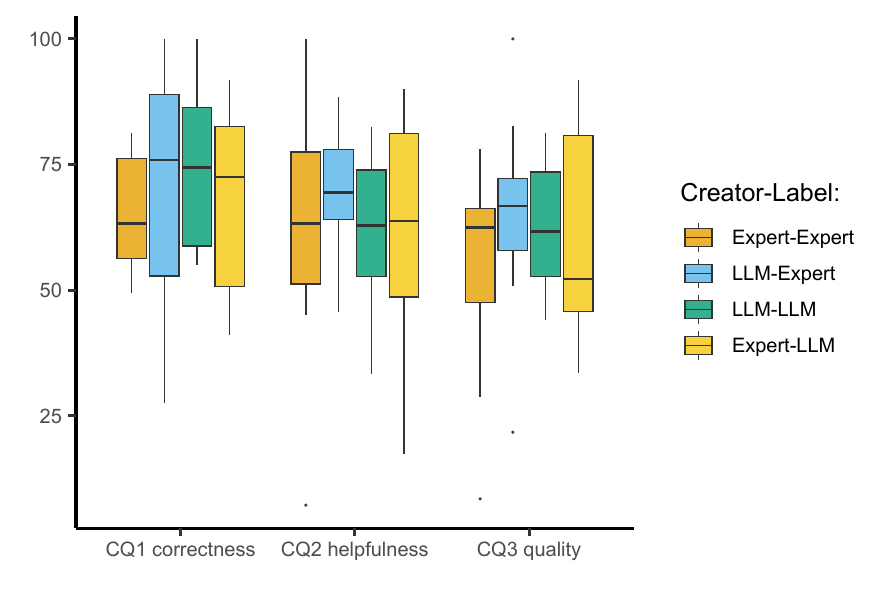}
    \caption{Boxplot with the scores of the custom questions (\Cref{tab:main_custom_questions}). The questions were rated on a visual analogue scale with values in a range from 0 to 100. The conditions are given as creator-label pairs.}
    \label{fig:soft-metrics-boxplot}
\end{figure}

\paragraph{Custom questions}
\label{sec:main_custom_questions}
We found no significant effects of the tip label or tip creator on our custom questions of correctness, helpfulness, or quality. A visualization of the custom questions can be seen in \Cref{fig:soft-metrics-boxplot}. In general, tips were of high quality ($\samplemean=61$, $\samplesd=19.18$), helpfulness ($\samplemean=64.41$, $\samplesd=18.66$) and correctness ($\samplemean=69.91$, $\samplesd=17.44$). 
Furthermore, there seems to be a tendency to rate questions as less difficult when the tip was labeled as \acrshort{llm}-generated ($\samplemean=40.2$, $\samplesd=17.9$) compared to expert-created ($\samplemean=49.9$, $\samplesd=16.8$).

\paragraph{Question difficulty}
\label{sec:question_difficulty}
The overall mean for the perceived question difficulty was $\samplemean=44.63$ ($\samplesd=17.91$), showing adequate difficulty levels for our question selection. Furthermore, we found that the simpler questions ($\samplemean=39.87$, $\samplesd=20.12$) where perceived as such, compared to the more advanced questions ($\samplemean=49.39$, $\samplesd=19.02$). Please refer to \Cref{sec:questions} for our selection process.

While we found no significant effect of tip label or creator on the perceived question difficulty, we found a tendency towards perceiving questions as less difficult when participants' were provided with a tip labeled as \acrshort{llm}-generated ($\samplemean=40.2$, $\samplesd=17.9$) compared to expert-labeled tips ($\samplemean=49.9$, $\samplesd=16.8$). There was no such tendency found for the tip creator, with only minimal differences between \acrshort{llm}-generated ($\samplemean=45.2$, $\samplesd=15.9$) and expert-created ($\samplemean=44.1$, $\samplesd=19.9$) tips.

\begin{figure}[htbp]
    \centering
    \begin{subfigure}{0.45\textwidth}
        \centering
        \includegraphics[height=5cm]{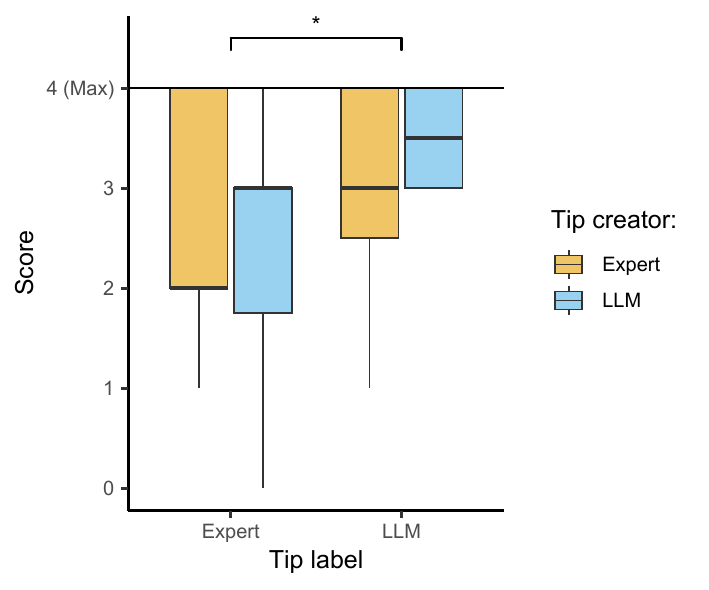}
    \end{subfigure}
    \hspace{0.05\textwidth}
    \begin{subfigure}{0.45\textwidth}
        \centering
        \includegraphics[height=5cm]{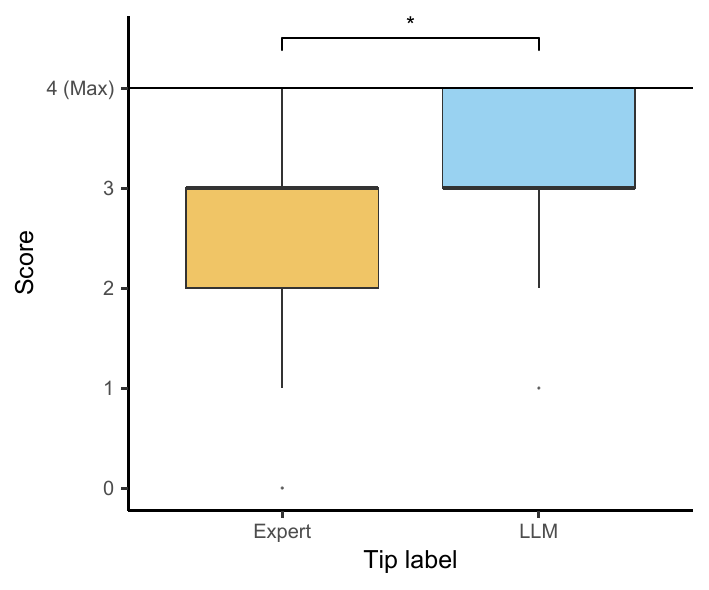}
    \end{subfigure}
    
    \caption{Two boxplots depicting the score achieved by main study participants dependent on creator and label of the provided tip (left) and the score achieved by main study participants dependent on only the label (right). Significant differences are marked with *.}
    \label{fig:score-labeled_as}
\end{figure}

\paragraph{Participant score}
\label{sec:main_store}
An analysis of our objective measure showed a significant effect (medium effect) for the participants score ($p<.05$, $\eta^2=.12$) for the factor tip label. The factor tip creator did not significantly affect the participant score. We can conclude that there is a positive influence on the score ($\samplemean_{Label:Expert}=2.48$, $\samplesd_{Label:Expert}=1.21$, $\samplemean_{Label:\acrshort{llm}}=3.28$, $\samplesd_{Label:\acrshort{llm}}=.84$), when the tip is labeled as \acrshort{llm}-generated (see \Cref{fig:score-labeled_as}). 


\subsection{Tip Evaluation Results}


Similarly to our main study, we conducted one-way ANOVAs using the factor tip creator. Note that this study only included one factor. Again, we constructed models for each of our measures (\Cref{tab:tip_eval_custom_questions}).
In cases where normality of the data was violated, we first align rank transformed the data using \acrshort{art}. Effect sizes are given using $\eta^2$ (Partial Eta Squared). We first evaluated if the previous study participation influenced how participants' rated the tips, but found no significant effect, omitting it for the analysis. The complete statistics are shown in \Cref{tab:tip_ratings}.

\begin{figure}[htb]
    \centering
    \includegraphics[width=.8\textwidth]{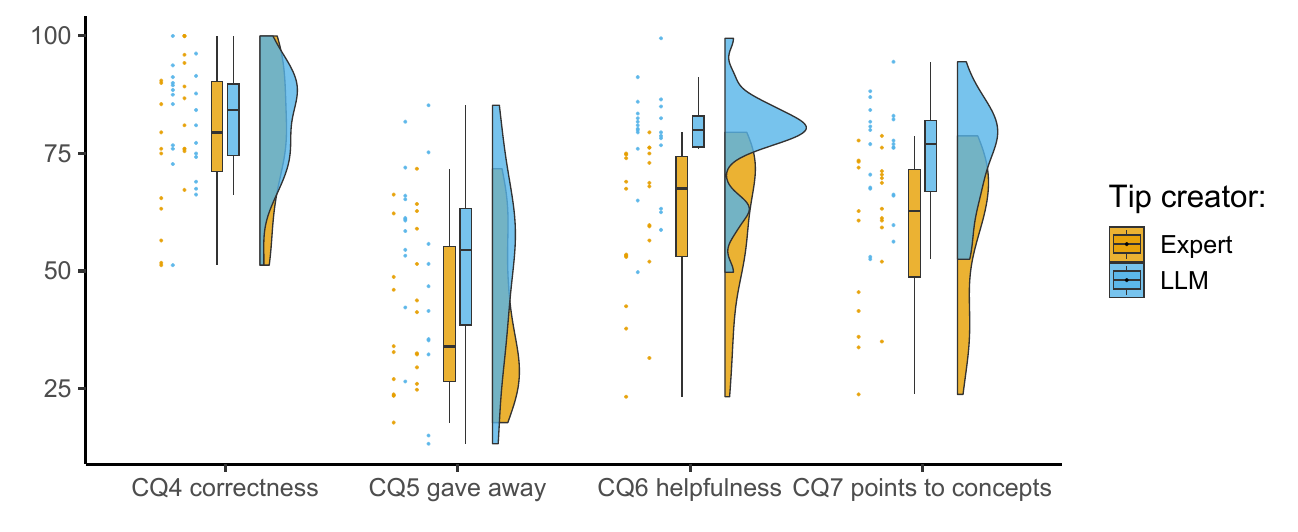}
    \caption{Raincloud plot showing the differences between \acrshort{llm}-generated and expert-created tips for the custom questions (\Cref{tab:tip_eval_custom_questions}). The raincloud plot consists of point clouds depicting the ratings given by participants', the boxplots and violin plots showing the distribution of ratings. All measures were rated on a visual analogue scale in a range from 0 to 100.}
    \label{fig:tip_eval_boxplot}
\end{figure}

\paragraph{Custom questions}
\label{sec:tip_eval_custom_questions}
We found no significant difference between \acrshort{llm}-generated and expert-created tips regarding the correctness. However, we found that \acrshort{llm}-generated tips scored significantly higher on giving away the answer($p<.05$, $\eta^2=.3$) compared to the expert-created tip. Furthermore, the \acrshort{llm}-generated tip was rated as significantly more helpful ($p<.0001$, $\eta^2=.51$) and as pointing significantly better towards relevant concepts ($p<.002$, $\eta^2=.36$) to answer the question. A visualization of these results is shown in \Cref{fig:tip_eval_boxplot}.

\begin{table*}[htb]
\centering
\caption{The ANOVA results of the tip evaluation study for the custom questions. Significance levels marked as ns := not significant; $p<0.05$ := *; $p<0.01$:=**; $p<0.001$:=***.}
\label{tab:tip_ratings}

\begin{tabular}{@{}lllllll@{}}
\toprule
\textbf{Measure} & \textbf{Df} & \textbf{Df.res} & \textbf{F value}&  \textbf{p-value} & \textbf{signif.} & \textbf{$\eta^2$ (partial)} \\
\midrule
correctness        & 1 & 22 & 0.15  & 0.702     & ns  & 6.77e-03\\
gave away          & 1 & 22 & 9.42  & 0.006     & **   & 0.30\\
helpfulness        & 1 & 22 & 22.81 & 9.075e-05 & *** & 0.51 \\
points to concepts & 1 & 22 & 12.42 & 0.002     & **  & 0.36\\
\bottomrule
\end{tabular}

\end{table*}

\begin{figure}[htb]
    \centering
    \includegraphics[height=5cm]{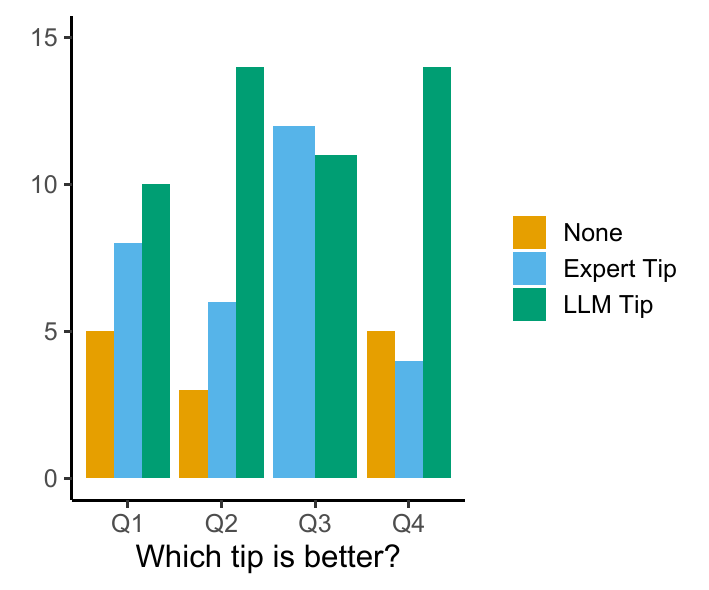}
    \caption{A histogram showing the amount of participants for each option when asked which tip was better for each question.}
    \label{fig:tip-rating-histogram}
\end{figure}

\paragraph{Direct tip comparison} We also asked participants to decide for each question, which tip they preferred (see \Cref{fig:tip-rating-histogram}). In three out of four questions, the participants rated the \acrshort{llm}-generated tip as better when directly compared to the expert-created tip. 

\begin{figure}[htbp]
    \centering
    \begin{subfigure}{0.22\textwidth}
        \centering
        \includegraphics[width=\textwidth]{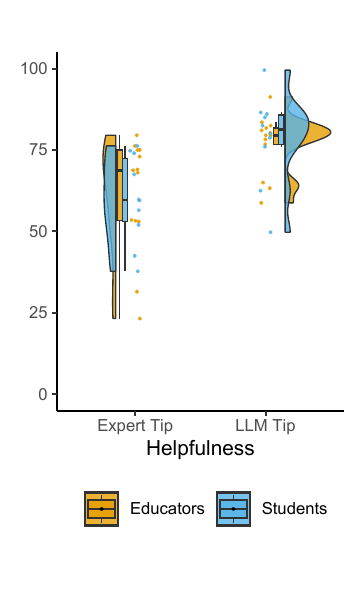}
    \end{subfigure}
    \hspace{0.01\textwidth}
    \begin{subfigure}{0.22\textwidth}
        \centering
        \includegraphics[width=\textwidth]{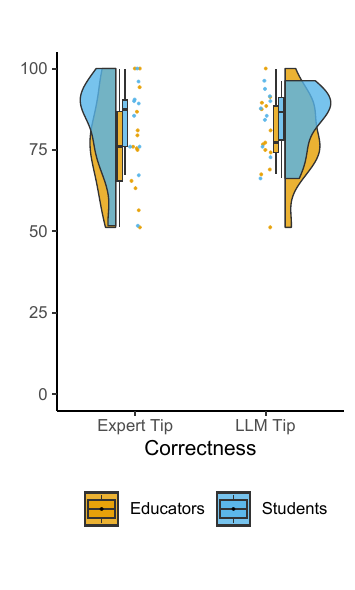}
    \end{subfigure}
    \hspace{0.01\textwidth}
    \begin{subfigure}{0.22\textwidth}
        \centering
        \includegraphics[width=\textwidth]{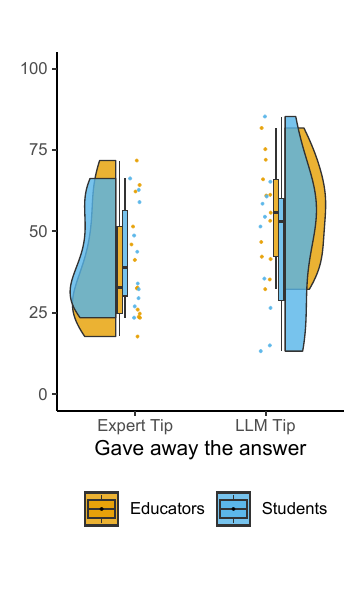}
    \end{subfigure}
    \hspace{0.01\textwidth}
    \begin{subfigure}{0.22\textwidth}
        \centering
        \includegraphics[width=\textwidth]{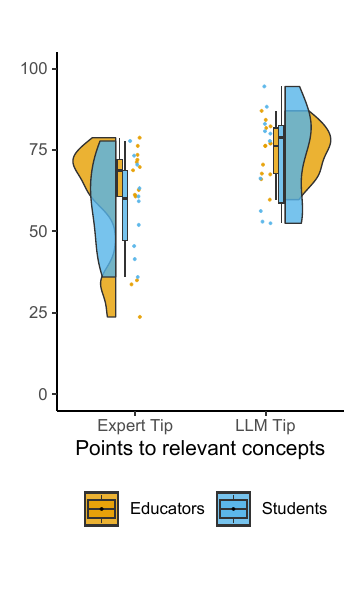}
    \end{subfigure}
    
    \caption{The raincloud plots for the custom questions of the tip evaluation study. The raincloud plot consists of point clouds depicting the ratings given by participants’, boxplots and violin plots showing the distribution of ratings. All measures were rated on a visual analogue scale in a range from 0 to 100.}
    \label{fig:tip-rating-raincloud}
\end{figure}

\paragraph{Differences between experts and students} When evaluating educators and students separately, we observed visual differences in the evaluation between the groups. As seen in \Cref{fig:tip-rating-raincloud}, both educators and students rate the helpfulness of \acrshort{llm}-generated tips more highly compared to expert-generated tips. Furthermore, both groups are in agreement for the rating of correctness and which tip better points to the relevant concepts. However, when looking at which tip gave away the answer more, the ratings of educators and students differ. While educators rate \acrshort{llm}-generated tips as giving more away, for students the ratings between \acrshort{llm}-generated tips and expert tips are relatively similar with only a slight tendency towards the \acrshort{llm}-generated tips giving away more.



    

\paragraph{Tip Length}
There are differences in the length of expert-created and \acrshort{llm}-generated tips. While the expert-created tips were generally short and on point, the \acrshort{llm}-generated tips tended to be lengthy and verbose. As one participant in the tip evaluation study put it, the \acrshort{llm}-generated tip \textit{"is more like a lecture note instead of a tip" (P20)}, highlighting the influence of the presentation on the perception of tips. In our qualitative evaluation (\Cref{sec:qual_analysis}), the increased length of \acrshort{llm}-generated tips was also an issue.

\begin{table}[htb]
    \centering
    \caption{The length of the tips (in words) given for each question in comparison.}
    \label{tab:tip_length}
    \begin{tabular}{ccccc|c}
    \toprule
    & Q1 & Q2 & Q3 & Q4 & Average \\
    \midrule
    Expert & 58 & 48 & 26 & 23 & 38.75 \\
    \acrshort{llm} & 110 & 148 & 162 & 182 & 150.5 \\
    \bottomrule
    \end{tabular}
    
\end{table}

\subsection{Qualitative Analysis}
\label{sec:qual_analysis}
In the tip evaluation study, we asked participants for their reasoning on which tip was better for each question. We encoded that reasoning in multiple steps. First, we reduced the arguments written by the participants to their key terms, which were then aggregated into six groups. We then counted the amount of occurrences for the expert tip and the \acrshort{llm} tip.


\begin{table*}[t]
\centering
\caption{The codings of the reasons given by participants' on why one of the tips is better then the other. It shows the amount of times the tips were rated that way.}
\label{tab:tip_codings}
\begin{tabular}{@{}lll@{}}
\toprule
\textbf{Group} & \textbf{\#Expert tip} &\textbf{\#LLM tip} \\
\midrule
bad content          & 4 & 7\\ 
points to concepts   & 4 & 8\\
adequate detail      & 3 & 8\\ 
adequate helpfulness & 9 & 35\\
gives away more      & 9 & 21\\ 
adequate length      & 3 & 3\\
\bottomrule
\end{tabular}

\end{table*}

The codings show that participants said more positive and more negative things about the \acrshort{llm}-generated tips. While participants mentioned a positive helpfulness 35 times for \acrshort{llm}-generated tips, they only mentioned it nine times for expert-created tips. Additionally, they mentioned that the \acrshort{llm}-generated tip contained good amount of detail ($n_{LLM}$=8, $n_{Expert}$=4) and points to relevant concepts ($n_{LLM}$=8, $n_{Expert}$=4) more often. However, they also mentioned that the \acrshort{llm}-generated tip gave away parts of the answer($n_{LLM}$=21, $n_{Expert}$=9) or produced bad content ($n_{LLM}$=7, $n_{Expert}$=4) more often.

\section{Discussion}\label{discussion}
Our results highlighted potential benefits of using \acrshort{llm}s for tip creation in the domain of quantum computing. In the following, we discuss these benefits and expose pitfalls that should be avoided.

\subsection{\acrshort{llm}-Generated Tips Rival Expert-Created Tips}

We found that participants in the tip evaluation study (\Cref{sec:tip_eval_custom_questions}) rated the helpfulness (CQ6) of \acrshort{llm}-generated tips significantly higher than that of expert-created tips. Yet, we have not found the same effect for helpfulness (CQ1) in our main study (\Cref{sec:main_custom_questions}), possibly due to the lack of direct comparison.
Participants in the main study might have rated it on how helpful the tip was to answer the question (CQ1), while participants of the tip evaluation study rated the helpfulness (CQ6) of the tips in comparison. As such, our results showcase that the helpfulness of \acrshort{llm}-generated tips is on par with expert-created tips and even prevail in a direct ranking (\textbf{RQ1b}). This is in line with other research about \acrshort{llm}-generated content~\cite{jung2023chatgpt}.

However, \acrshort{llm}-generated tips were more prone to give away the correct answer (CQ5), possibly correlating with their perceived helpfulness (CQ6) and them pointing to relevant concepts (CQ7). Yet, this caveat was only present for the educator group in our tip evaluation study (\Cref{fig:tip-rating-raincloud}). We believe that their different perception stems from the educators' teaching experience which the students are lacking. Consequently, there is a real possibility that \acrshort{llm}-generated tips more easily give away the correct answer (\textbf{RQ2}). This is a pitfall that needs to be avoided through proper prompt engineering~\cite{lee2023few} or possibly using an \acrshort{llm} as a judge~\cite{zheng2024judging} that is specifically trained to evaluate whether the \acrshort{llm}-generated tip gives away the answer.


We found no significant differences between the correctness of \acrshort{llm}-generated tips and expert-created tips in the main study (CQ3) or in the tip evaluation study (CQ4). Participants rated all tips --- regardless of creator or label --- as correct (\Cref{sec:main_custom_questions}) contributing to answering \textbf{RQ1b}. 

Finally, the \acrshort{llm}-generated tips pointed to relevant concepts (CQ7) significantly better compared to the expert-created tips, a characteristic that is highly valuable in teaching~\cite{krathwohl2002revision, erickson2002concept}.

In addition to the positive subjective perception of \acrshort{llm}-generated tips, students did not score worse in our multiple-choice questions (main study, see \Cref{sec:main_study}). Here, \acrshort{llm}-generated tips did not negatively impact students' scores (\Cref{sec:main_store}), answering \textbf{RQ1a}. 


Our results show that \acrshort{llm}s can be used successfully to generate tips in the context of basic quantum computing education, rivaling expert-created tips in all metrics, answering our first research question (\textbf{RQ1}). 

\subsection{Novelty of \acrshort{ai} Possibly Leads to Increased Scrutiny}
While the tip creator did not influence the students' scores, a tip labeled as being \acrshort{llm}-generated (regardless of its original creator) elicited significantly higher scores.
We believe that this phenomenon occurred due to an increase of attention given to tips labeled as \acrshort{llm}-generated. Due to the exhaustive media attention given to \acrshort{llm}s since the inception of ChatGPT, participants came into contact with these models. While some might have only heard of them, others might have tried them out or may even be actively using \acrshort{llm}s for some tasks. 
This knowledge could have made them aware of the persisting issues attributed to this technology. Namely hallucinations and confidently worded verbose texts lacking substance~\cite{tonmoy2024comprehensive}. As a consequence, participants who were told that they were given an \acrshort{llm}-generated tip, might have examined the tip more carefully, thinking critically about its contents. This critical thinking might have led to a better understanding and, therefore, an improved score~\cite{pithers2000critical}. The measurable impact of introducing \acrshort{llm} labels is an important design consideration to make and needs to be handled carefully (\textbf{RQ2}).



We envision this effect due to two interacting factors impacting the students' perceptions of the offered support (the \acrshort{ai}-generated tip) and the resulting perceived difficulty of the task.

Firstly, a measurable placebo effect has been shown to exist in previous work on human-computer interaction~\cite{villa2023placebo, kosch2023placebo,kloft2024ai}. While most works were only able to find this placebo effect for subjective metrics, Kloft et al.~\cite{kloft2024ai} observed a significant objective performance improvement in a study when comparing participants using an active sham \acrshort{ai} with participants using an inactive sham \acrshort{ai}. As such, \acrshort{ai}-enabled support is seen as something desirable.

Secondly, studies in various contexts have found an impact of perception of difficulty on test performance~\cite{HONG1999431,KUKLA1974374,sarrazin2002exerted}. High difficulty leads to dissatisfaction, while perceived low difficulty can be motivating. Labeling a tip as \acrshort{llm}-generated actually had the impact of lowering the perceived difficulty of the question (see \Cref{sec:question_difficulty}) compared to labeling a tip as expert-created, possibly reducing reluctance and increasing motivation, resulting in increased performance.


\subsection{Opportunities and Challenges of LLMs for Teaching}

While our work has revealed design pitfalls to avoid when working with \acrshort{llm}s, such as a possible placebo effect as well as lengthy and leading tips (\textbf{RQ2}), there is a real opportunity to use \acrshort{llm}s for generating content about the basics of quantum computing. In general, \acrshort{llm}-generated tips have performed equally well, if not better, than their expert-created counterparts. Especially in a direct comparisons, \acrshort{llm}-generated tips were favored and conveyed the concepts for students to understand and learn better that expert-created tips.

Whilst supporting educators for tip creation, we believe that \acrshort{llm}s can shine specifically when used to individualize content for education. Recent works have already successfully used \acrshort{llm}s to generate math questions and answers based on a students interests~\cite{christ2024mathwell} through prompt engineering~\cite{lee2023few}.
It can be a powerful tool strongly dictating the quality of the generated content. If done right, prompt engineering can solve some of the design caveats of \acrshort{llm}-generated tips we found in our work, such as reducing the generated tip length and minimizing the occurrence of leading tips.


Ultimately, there will always be a trade-off with respect to correctness and quality of \acrshort{llm}-generated tips. 
We made sure that the \acrshort{llm}-generated tips in our study were correct, which was also confirmed by participants. However, there remains a risk that \acrshort{llm}-generated tips might be incorrect or misleading, especially for more complex questions~\cite{krupp2024unreflected}.
Although we can reduce this risk, there is no way to ensure that each tip an \acrshort{llm} generates will be correct. Consequently, the need for formal validation of \acrshort{llm}-produced content arises. While methods to validate \acrshort{ai} models exist, validating them is a complex task requiring novel methods~\cite{huang2024survey}.

Especially in the domain of education, misinformation and incorrect content are detrimental for the learner's progress~\cite{holmes2013data}. As such, we advice for a step-by-step integration of \acrshort{llm}-generated content for education, starting with human-in-the-loop evaluations~\cite{amirizaniani2024developing}, applications in low-risk scenarios (such as tips) and potentially applying \acrshort{llm} judges~\cite{zheng2024judging} as a secondary safety net.



\subsection{Limitations}
Our results are limited to our study setup. They only show that \acrshort{llm}-generated tips can be used instead of expert-created tips for basic quantum computing exercises using GPT4 with a \acrshort{rag}-like approach to tip-creation. The results may change for other \acrshort{llm}s, creating tips for different domains, or increasing the question complexity. Before applying this method in practice, experts should ensure that the generated tips are of adequate quality for the questions at hand. Even using a \acrshort{rag}-like approach, the \acrshort{llm} might hallucinate, leading to wrong or misleading tips. This underlines the importance of quality control and formal verification.



Ensuring an equal level of quantum computing knowledge for our participants was challenging. In this work, we used the QUIKSTART 2024 winterschool to ensure that all participants had adequate knowledge of the topic and also used it as distribution platform, ensuring a consistent experience level among participants. Bigger cohorts can potentially be achieved by integrating the study within dedicated study programs.


\section{Conclusion}\label{conclusion}

\acrshort{llm}-generated content has permeated almost every area in our everyday lives. In this work, we investigated the potential of \acrshort{llm}s to generate adequate educational content. In particular, we focused on how \acrshort{llm}s can support --- not substitute --- educators in domains with low educator-to-student ratios. Our results are promising, highlighting that \acrshort{llm}s are capable of producing high quality tips on par with expert-created tips, albeit are subject to design considerations including a possible placebo effect of \acrshort{ai} and a tendency to generate leading tips. We discuss how these findings can support teaching methods pointing towards \acrshort{llm}-supported smart education, and expose challenges still to overcome when employing \acrshort{llm}-generated content.


\printglossary[type=\acronymtype, title={List of abbreviations}]

\backmatter





\bmhead{Acknowledgements}

The authors thank all the study participants and all guest speakers of the QUIP quantum machine learning winter school QUIKSTART 2024: Antonio Macaluso, Nico Piatkowski, Carsten Blank, Bob Coecke, Shai Machnes, Iraitz Montalban, Caitlin Jones, and Sebastian Stern. Also, we are grateful to Antonio Macaluso, Christina Jörg, Alexey Bochkarev, Nikolaos Palaiodimopoulos, and Thomas Niederprüm for their additional expertise on the topic. Finally, we would like to thank the helpers beyond the scientific program, especially Verena Paff, Jörn Gukenbiehl and Luis Sanchez. 

A.W., H.O., P.L, V.B. and M.K-E. acknowledge support from the Quantum Initiative Rhineland-Palatinate (QUIP). A.W., H.O., P.L, and M.K-E. acknowledge support from the Research Initiative Quantum Computing for Artificial Intelligence (QC-AI).

J.B., A.W., P.L., and M.K-E. acknowledge support by the QuanTUK project at the RPTU in Kaiserslautern, supported by the Federal Ministry of Education and Research (FKZ13N15995).

L.K., J.K., and P.L. acknowledge support by the KI4TUK project at the RPTU in Kaiserslautern, supported by the Federal Ministry of Education and Research (BMBF) under grant number 16DHBKI058.

\section*{Declarations}

\bmhead{Availability of data and materials} The datasets generated and analyzed during the current study are available in the following repository: \url{https://doi.org/10.5281/zenodo.12723463}.

\bmhead{Competing interests} The authors declare that they have no conflict of interest.

\bmhead{ Ethics approval was granted by the Ethics Department of the German Research Center for Artifical Intelligence (DFKI)}

\bmhead{Authors' contributions} 
The main study was performed during the QUIP winter school QUIKSTART 2024 at RPTU in Kaiserslautern Germany and was mainly organized by V.B. and M.K-E.. QUIKSTART 2024 was mainly possible due to contributed lectures, expertise and supervision of
I.G., A.G., S.B., S.S., A.M., A.C.M., A.W., H.O., and P.L..

L.K., and J.K. designed the studies, J.B. and M.K-E. added quantum computing related questions. L.K., J.B., J.K. and M.K-E. finalized the design. I.G. and A.G. provided expert tips and presented the corresponding elementary lectures on quantum computing. S.M., and S.S. provided lectures on advanced quantum and machine learning. A.M., A.C.M, P.L., and V.B. supervised the elementary and advanced quantum computing, machine learning lectures and scripts.
L.K., J.B., I.G., A.G., V.B. and M.K.-E. collected the data. 
L.K. and J.K. evaluated and analyzed the data.
L.K., J.B., J.K., and M.K-E. discussed the data. L.K., J.B., J.K., and M.K.-E. wrote the first draft of the manuscript. P.L., and J.K. supervised the study. All authors have read, reviewed and approved the final manuscript.









\begin{appendices}

\section{Multiple Choice Questions}\label{secA1}
\subsection{Question 1}
What is the following state on the Bloch sphere?

$|\psi\rangle = \frac{1}{\sqrt{2}} |0\rangle + \frac{1}{\sqrt{2}} |1\rangle$

\begin{figure}[h!]
    \centering
    \begin{tabular}{cc}
        \begin{subfigure}[b]{0.45\textwidth}
            \centering
            \includegraphics[width=\textwidth]{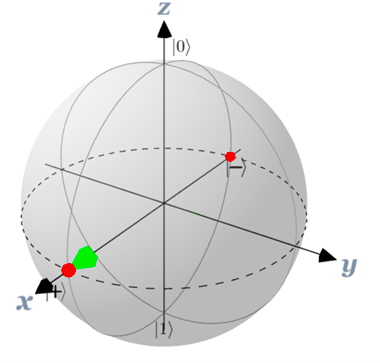}
            \caption{correct}
            \label{fig:1}
        \end{subfigure} &
        \begin{subfigure}[b]{0.45\textwidth}
            \centering
            \includegraphics[width=\textwidth]{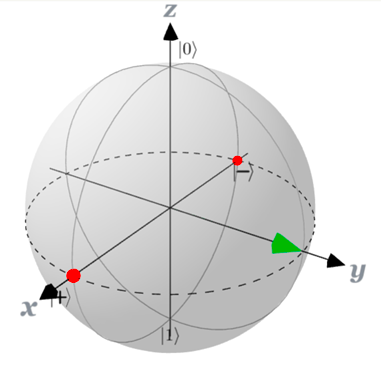}
            \caption{incorrect}
            \label{fig:2}
        \end{subfigure} \\
        \begin{subfigure}[b]{0.45\textwidth}
            \centering
            \includegraphics[width=\textwidth]{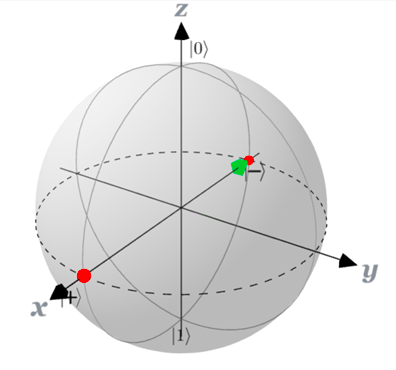}
            \caption{incorrect}
            \label{fig:3}
        \end{subfigure} &
        \begin{subfigure}[b]{0.45\textwidth}
            \centering
            \includegraphics[width=\textwidth]{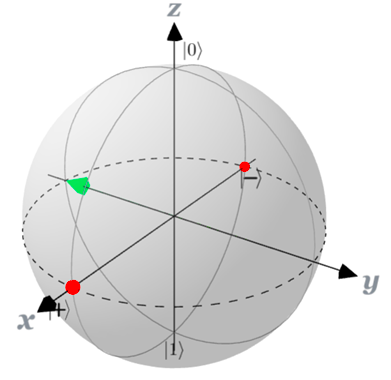}
            \caption{incorrect}
            \label{fig:4}
        \end{subfigure}
    \end{tabular}
    \caption{Possible answer options for question 1.}
    \label{appendix:q1}
\end{figure}

\subsection{Question 2}
What is the following state on the Bloch sphere?
$|\psi\rangle = \frac{1}{2} |0\rangle + \frac{\sqrt{3}}{2} e^{i\frac{\pi}{2}} |1\rangle$
\begin{figure}[h!]
    \centering
    \begin{tabular}{cc}
        \begin{subfigure}[b]{0.45\textwidth}
            \centering
            \includegraphics[width=\textwidth]{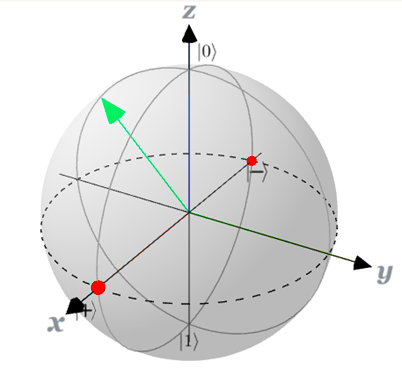}
            \caption{incorrect}
            \label{fig:1}
        \end{subfigure} &
        \begin{subfigure}[b]{0.45\textwidth}
            \centering
            \includegraphics[width=\textwidth]{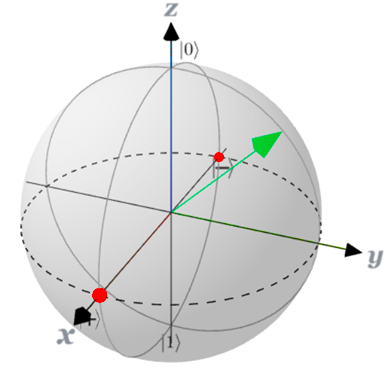}
            \caption{incorrect}
            \label{fig:2}
        \end{subfigure} \\
        \begin{subfigure}[b]{0.45\textwidth}
            \centering
            \includegraphics[width=\textwidth]{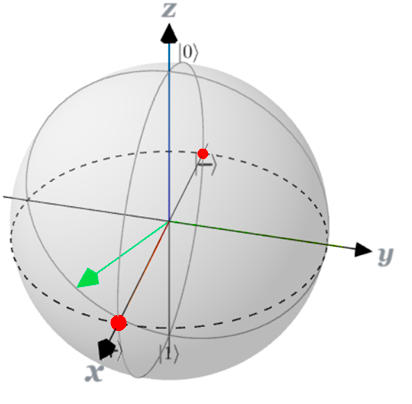}
            \caption{incorrect}
            \label{fig:3}
        \end{subfigure} &
        \begin{subfigure}[b]{0.45\textwidth}
            \centering
            \includegraphics[width=\textwidth]{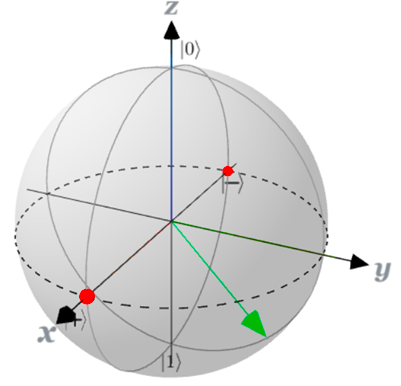}
            \caption{correct}
            \label{fig:4}
        \end{subfigure}
    \end{tabular}
    \caption{Possible answer options for question 2.}
    \label{appendix:q2}
\end{figure}

\subsection{Question 3}
What is equal to the following two-qubit operation:\\
(H $\otimes$ I) CNOT CNOT (H $\otimes$ I)?\\
\begin{figure}
    \centering
    \includegraphics{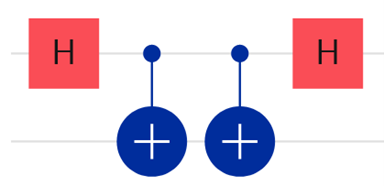}
    \caption{Visualization of question 3.}
    \label{appendix:q3}
\end{figure}
a) I$\otimes$I (correct)\\
b) H$\otimes$I\\
c) CNOT\\
d) (H$\otimes$I) CNOT\\
\subsection{Question 4}
Which Bell-state is the outcome of the following operation on the $|00$⟩ state:\\
(X $\otimes$ I) CNOT((H $\otimes$ I)$|00$⟩)?\\
\begin{figure}
    \centering
    \includegraphics{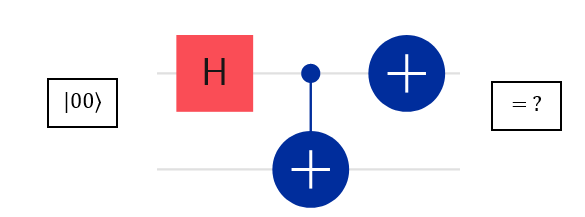}
    \caption{Visualization of question 4.}
    \label{appendix:q4}
\end{figure}
a) $\frac{1}{\sqrt{2}} \left( |00\rangle + |11\rangle \right)$\\
b) $\frac{1}{\sqrt{2}} \left( |00\rangle - |11\rangle \right)$\\
c) $\frac{1}{\sqrt{2}} \left( |01\rangle + |10\rangle \right)$ (correct)\\
d) $\frac{1}{\sqrt{2}} \left( |01\rangle - |10\rangle \right)$\\




\section{Tips}
\subsection{Question 1}

\paragraph{Expert Tip:}
Realize you are in an equal superposition: Compute the probability of outcome 0 and 1. On the Bloch sphere an outcome is probable if the state vector is closer to the pole corresponding to the outcome. The definition of the relative phase, relative to the x-axis on the Bloch sphere, gives you the representation on the Bloch sphere.

\paragraph{LLM Tip:}
To understand where the given state $|\psi\rangle = \frac{1}{\sqrt{2}} |0\rangle + \frac{1}{\sqrt{2}} |1\rangle$ is on the Bloch sphere, remember that each point on the Bloch sphere represents a possible quantum state of a qubit. The state vector $|0\rangle$ corresponds to the north pole of the Bloch sphere, and $|1\rangle$ corresponds to the south pole. A state like $| \psi \rangle$ which is an equal superposition of $|0\rangle$ and $|1\rangle$ with no complex phase (meaning both coefficients are real and positive) will be on the equator of the Bloch sphere. Look for a state vector that is equally between $|0\rangle$ and $|1\rangle$ on the equator, without any phase difference indicated by complex numbers.

\subsection{Question 2}
\paragraph{Expert Tip:}
On the Bloch sphere, the amplitude of state $|0\rangle$ is related to the sine of $\vartheta$, the polar angle. The amplitude of $|1\rangle$ is related to the cosine of $\vartheta$ but one also needs to account for the relative phase, which is represented by the azimuthal angle $\varphi$.

\paragraph{LLM Tip:}
When trying to identify a state on the Bloch sphere, you'll want to consider both the magnitude and the phase of the qubit state coefficients. For the state $\psi\rangle = \frac{1}{2} |0\rangle + \frac{\sqrt{3}}{2} e^{i\frac{\pi}{2}} |1\rangle$, note the coefficients in front of $|0\rangle$ and $|1\rangle$ . The magnitude squared of these coefficients will give you the probabilities of measuring each state. The phase factor $e^{i\frac{\pi}{2}}$ corresponds to a phase of $\frac{\pi}{2}$ radians or 90 degrees. On the Bloch sphere, states are represented as points on the surface of the sphere, where the angles theta ($\theta$) and phi ($\phi$) determine the position. Theta is the angle from the positive z-axis and phi is the angle from the positive x-axis in the x-y plane. To find the correct picture, look for the one where the vector representing the qubit state has these properties.

\subsection{Question 3}
\paragraph{Expert Tip:}
Many quantum gates are their own inverse. Think of the outcome of applying the CX-gate twice. Think of the outcome of applying the Hadamard gate twice.

\paragraph{LLM Tip:}
Let's break down the operation given in your question, step by step, using concepts from the document. The Hadamard gate (H) puts a qubit into a superposition state, meaning it creates a combination of all possible states. The Identity gate (I) does nothing to the qubit it acts upon. The CNOT gate is a conditional gate that flips the target qubit if the control qubit is in the state $|1\rangle$.
 
Here's a hint to think about how each component of the operation affects the state of the qubits:
\begin{itemize}
    \item Start by understanding the effect of the Hadamard gate on a single qubit and how the Identity gate leaves the other qubit unchanged.
    \item Then, consider what happens when a CNOT gate acts on two qubits, specifically how it changes the state of the qubits based on the condition mentioned above.
    \item Finally, think about the sequence of operations and their reversibility, especially how applying the same operations in reverse order might affect the qubits' state.
\end{itemize}
    
\subsection{Question 4}
\paragraph{Expert Tip:}
Recall the bell state generator (Hadamard + CX-gate). The additional X-gate in the circuit changes the state of one qubit in the outcome.

\paragraph{LLM Tip:}
To understand which Bell state is produced by the given quantum circuit, let's break down the operation step by step:
\begin{itemize}
    \item The Hadamard gate ($H\otimes I$) acts on the $|00\rangle$ state. The Hadamard gate puts the first qubit into a superposition state, transforming $|0\rangle$ into $\frac{|0⟩ + |1\rangle}{\sqrt{2}}$ .
    \item The CNOT gate acts on this new state. The CNOT gate is a conditional gate that flips the second qubit if the first qubit is in the $|1\rangle$ state.
    \item Finally, the Pauli-X gate ($X\otimes I$) acts only on the first qubit, flipping it from $|0\rangle$ to $|1\rangle$ and vice versa.
\end{itemize}

Your task is to apply these operations successively, keeping in mind the effects of each gate. The Hadamard gate creates a superposition, the CNOT entangles qubits by flipping the second based on the first, and the X gate flips the qubit from $|0\rangle$ to $|1\rangle$ or $|1\rangle$ to $|0\rangle$. The final state will be a Bell state, an entangled state of two qubits. Look at the transformation of each qubit at each step, and this will guide you to the correct Bell state.

\section{Tip Generation}
We used the following prompt to generate our tips with GPT-4.
\begin{quote}
    You are a passionate and patient quantum-physics teacher. Based on the information given in the pdf, give the struggling student a hint that helps him answer the following question, without giving away the answer. Be more specific and formulate your tip as if you were talking to a novice. Be as concise as possible.
    
    \{Question + answers + correct answer\}
\end{quote}
Also adding the pdf of the quantum computing basics used in QUICKSTART 2024 and the images of the questions as context. 
\section{Additional Results}
\label{additional_results}
\begin{table}[ht]
    \centering
    \caption{The ANOVA results of the main study for the participants' score.}
    \label{tab:main_test_results_score}
    \begin{tabular}{llllllll}
        \toprule
        \textbf{Measure} & \textbf{Factor} &  Df & Df.res & F value &  p-value & signif. & $\eta^2$ (partial)\\
                 \midrule
              score &    tip creator & 1 & 42 &  0.34 &  0.56 & ns & 8.06e-03\\   
              score &    tip label & 1 & 42 &  5.89 & 0.02 & * & 0.12\\
            \bottomrule
    \end{tabular}
    
\end{table}

\begin{table}[ht]
    \centering
    \caption{The ANOVA results of the main survey for the perceived question difficulty.}
    \label{tab:main_test_results_difficulty}
    \begin{tabular}{llllllll}
        \toprule
        \textbf{Measure} & \textbf{Factor} &  Df & Df.res & F value &  p-value & signif. & $\eta^2$ (partial)\\
                 \midrule
          difficulty &    tip creator & 1 & 42 &  0.14 & 0.71 & ns & 3.44e-03\\
           difficulty &   tip label & 1 & 42 & 3.08 & 0.09 & ns & 0.07\\
            \bottomrule
    \end{tabular}
    
\end{table}

\begin{table}[ht]
    \centering
    \caption{The ANOVA results of the main survey for the custom questions.}
    \label{tab:main_test_results_custom_questions}
    \begin{tabular}{llllllll}
        \toprule
        \textbf{Measure} & \textbf{Factor} &  Df & Df.res & F value &  p-value & signif. & $\eta^2$ (partial)\\
              \midrule
          correctness &    tip creator & 1 & 42 &  1.60 & 0.21 & ns & 0.04 \\    
          correctness &    tip label & 1 & 42 &  0.22 & 0.64 & ns & 5.14e-03\\
          \midrule
          helpfulness &    tip creator & 1 & 42 &  0.22 & 0.64 & ns & 5.30e-03\\ 
          helpfulness &    tip label & 1 & 42 &  0.43 & 0.52 & ns & 0.01\\
          \midrule
          quality &    tip creator  & 1 & 42 &  0.58 & 0.45& ns & 0.01\\
          quality &    tip label   & 1 & 42 &  0.03 & 0.86 & ns & 7.81e-04\\

            \bottomrule
    \end{tabular}
    
\end{table}

\begin{table}[ht]
    \centering
    \caption{The ANOVA results of the main survey for the custom questions.}
    \label{tab:main_test_results_custom_questions}
    \begin{tabular}{llllllll}
        \toprule
        \textbf{Measure} & \textbf{Factor} &  Df & Df.res & F value &  p-value & signif. & $\eta^2$ (partial)\\
              \midrule
          difficulty &    tip creator & 1 & 42 &  0.14 &0.71 & ns & 3.44e-03 \\    
          difficulty &    tip label & 1 & 42 &  3.08 & 0.09 & ns & 0.07\\
            \bottomrule
    \end{tabular}
    
\end{table}

\end{appendices}
\FloatBarrier

\bibliography{bibliography}

\end{document}

%% file: glossary.tex
\makeglossaries

\newacronym{cf}{CF}{Competence Framework}

\newacronym{qist}{QIST}{Quantum Information Science \& Technology}

\newacronym{ecl}{ECL}{Extraneous Cognitive Load}

\newacronym{gcl}{GCL}{Germane Cognitive Load}

\newacronym{cn}{CN}{Circle Notation}

\newacronym{dcn}{DCN}{Dimensional Circle Notation}

\newacronym{icl}{ICL}{Intrinsic Cognitive Load}

\newacronym{deft}{DeFT}{Design, Functions, Tasks}

\newacronym{mer}{MERs}{multiple external representations}

\newacronym{clt}{CLT}{Cognitive Load Theory}

\newacronym{stem}{STEM}{Science, Technology, Engineering \& Mathematics}

\newacronym{llm}{LLM}{Large Language Model}

\newacronym{quip}{QUIP}{Quantum-Initiative Rhineland-Palatinate}

\newacronym{rag}{RAG}{Retrieval-Augmented Generation}

\newacronym{bf}{BF}{Bayes Factor}

\newacronym{ai}{AI}{Artificial Intelligence}

\newacronym{qc}{QC}{Quantum Computing}

\newacronym{art}{ART}{Aligned Rank Transform}

\newacronym{anova}{ANOVA}{Analysis of Variance}

\newacronym{dfki}{DFKI}{German Research Center for Artificial Intelligence}

%% file: main.bbl

\begin{thebibliography}{79}
\ifx \bisbn   \undefined \def \bisbn  #1{ISBN #1}\fi
\ifx \binits  \undefined \def \binits#1{#1}\fi
\ifx \bauthor  \undefined \def \bauthor#1{#1}\fi
\ifx \batitle  \undefined \def \batitle#1{#1}\fi
\ifx \bjtitle  \undefined \def \bjtitle#1{#1}\fi
\ifx \bvolume  \undefined \def \bvolume#1{\textbf{#1}}\fi
\ifx \byear  \undefined \def \byear#1{#1}\fi
\ifx \bissue  \undefined \def \bissue#1{#1}\fi
\ifx \bfpage  \undefined \def \bfpage#1{#1}\fi
\ifx \blpage  \undefined \def \blpage #1{#1}\fi
\ifx \burl  \undefined \def \burl#1{\textsf{#1}}\fi
\ifx \doiurl  \undefined \def \doiurl#1{\url{https://doi.org/#1}}\fi
\ifx \betal  \undefined \def \betal{\textit{et al.}}\fi
\ifx \binstitute  \undefined \def \binstitute#1{#1}\fi
\ifx \binstitutionaled  \undefined \def \binstitutionaled#1{#1}\fi
\ifx \bctitle  \undefined \def \bctitle#1{#1}\fi
\ifx \beditor  \undefined \def \beditor#1{#1}\fi
\ifx \bpublisher  \undefined \def \bpublisher#1{#1}\fi
\ifx \bbtitle  \undefined \def \bbtitle#1{#1}\fi
\ifx \bedition  \undefined \def \bedition#1{#1}\fi
\ifx \bseriesno  \undefined \def \bseriesno#1{#1}\fi
\ifx \blocation  \undefined \def \blocation#1{#1}\fi
\ifx \bsertitle  \undefined \def \bsertitle#1{#1}\fi
\ifx \bsnm \undefined \def \bsnm#1{#1}\fi
\ifx \bsuffix \undefined \def \bsuffix#1{#1}\fi
\ifx \bparticle \undefined \def \bparticle#1{#1}\fi
\ifx \barticle \undefined \def \barticle#1{#1}\fi
\bibcommenthead
\ifx \bconfdate \undefined \def \bconfdate #1{#1}\fi
\ifx \botherref \undefined \def \botherref #1{#1}\fi
\ifx \url \undefined \def \url#1{\textsf{#1}}\fi
\ifx \bchapter \undefined \def \bchapter#1{#1}\fi
\ifx \bbook \undefined \def \bbook#1{#1}\fi
\ifx \bcomment \undefined \def \bcomment#1{#1}\fi
\ifx \oauthor \undefined \def \oauthor#1{#1}\fi
\ifx \citeauthoryear \undefined \def \citeauthoryear#1{#1}\fi
\ifx \endbibitem  \undefined \def \endbibitem {}\fi
\ifx \bconflocation  \undefined \def \bconflocation#1{#1}\fi
\ifx \arxivurl  \undefined \def \arxivurl#1{\textsf{#1}}\fi
\csname PreBibitemsHook\endcsname

\bibitem[\protect\citeauthoryear{MacFarlane et~al.}{2003}]{MacFarlane2003}
\begin{barticle}
\bauthor{\bsnm{MacFarlane}, \binits{A.G.J.}},
\bauthor{\bsnm{Dowling}, \binits{J.P.}},
\bauthor{\bsnm{Milburn}, \binits{G.J.}}:
\batitle{Quantum technology: the second quantum revolution}.
\bjtitle{Philosophical Transactions of the Royal Society of London. Series A: Mathematical, Physical and Engineering Sciences}
\bvolume{361}(\bissue{1809}),
\bfpage{1655}--\blpage{1674}
(\byear{2003})
\doiurl{10.1098/rsta.2003.1227}
{\href{https://arxiv.org/abs/https://royalsocietypublishing.org/doi/pdf/10.1098/rsta.2003.1227}{{https://royalsocietypublishing.org/doi/pdf/10.1098/rsta.2003.1227}}}
\end{barticle}
\endbibitem

\bibitem[\protect\citeauthoryear{Roberson}{2021}]{Roberson2021}
\begin{barticle}
\bauthor{\bsnm{Roberson}, \binits{T.M.}}:
\batitle{On the social shaping of quantum technologies: An analysis of emerging expectations through grant proposals from 2002--2020}.
\bjtitle{Minerva}
\bvolume{59}(\bissue{3}),
\bfpage{379}--\blpage{397}
(\byear{2021})
\doiurl{10.1007/s11024-021-09438-5}
\end{barticle}
\endbibitem

\bibitem[\protect\citeauthoryear{Scheidsteger et~al.}{2022}]{Scheidsteger2022}
\begin{barticle}
\bauthor{\bsnm{Scheidsteger}, \binits{T.}},
\bauthor{\bsnm{Haunschild}, \binits{R.}},
\bauthor{\bsnm{Ettl}, \binits{C.}}:
\batitle{Historical roots and seminal papers of quantum technology 2.0}.
\bjtitle{NanoEthics}
\bvolume{16}(\bissue{3}),
\bfpage{271}--\blpage{296}
(\byear{2022})
\doiurl{10.1007/s11569-022-00424-z}
\end{barticle}
\endbibitem

\bibitem[\protect\citeauthoryear{Roberson et~al.}{2023}]{Roberson_Raman_Leach_Vilkins_2023}
\begin{barticle}
\bauthor{\bsnm{Roberson}, \binits{T.}},
\bauthor{\bsnm{Raman}, \binits{S.}},
\bauthor{\bsnm{Leach}, \binits{J.}},
\bauthor{\bsnm{Vilkins}, \binits{S.}}:
\batitle{Assessing the journey of technology hype in the field of quantum technology}.
\bjtitle{TATuP - Zeitschrift für Technikfolgenabschätzung in Theorie und Praxis}
\bvolume{32}(\bissue{3}),
\bfpage{17}--\blpage{21}
(\byear{2023})
\doiurl{10.14512/tatup.32.3.17}
\end{barticle}
\endbibitem

\bibitem[\protect\citeauthoryear{Arute et~al.}{2019}]{Arute2019}
\begin{barticle}
\bauthor{\bsnm{Arute}, \binits{F.}},
\bauthor{\bsnm{Arya}, \binits{K.}},
\bauthor{\bsnm{Babbush}, \binits{R.}},
\bauthor{\bsnm{Bacon}, \binits{D.}},
\bauthor{\bsnm{Bardin}, \binits{J.C.}},
\bauthor{\bsnm{Barends}, \binits{R.}},
\bauthor{\bsnm{Biswas}, \binits{R.}},
\bauthor{\bsnm{Boixo}, \binits{S.}},
\bauthor{\bsnm{Brandao}, \binits{F.G.S.L.}},
\bauthor{\bsnm{Buell}, \binits{D.A.}},
\bauthor{\bsnm{Burkett}, \binits{B.}},
\bauthor{\bsnm{Chen}, \binits{Y.}},
\bauthor{\bsnm{Chen}, \binits{Z.}},
\bauthor{\bsnm{Chiaro}, \binits{B.}},
\bauthor{\bsnm{Collins}, \binits{R.}},
\bauthor{\bsnm{Courtney}, \binits{W.}},
\bauthor{\bsnm{Dunsworth}, \binits{A.}},
\bauthor{\bsnm{Farhi}, \binits{E.}},
\bauthor{\bsnm{Foxen}, \binits{B.}},
\bauthor{\bsnm{Fowler}, \binits{A.}},
\bauthor{\bsnm{Gidney}, \binits{C.}},
\bauthor{\bsnm{Giustina}, \binits{M.}},
\bauthor{\bsnm{Graff}, \binits{R.}},
\bauthor{\bsnm{Guerin}, \binits{K.}},
\bauthor{\bsnm{Habegger}, \binits{S.}},
\bauthor{\bsnm{Harrigan}, \binits{M.P.}},
\bauthor{\bsnm{Hartmann}, \binits{M.J.}},
\bauthor{\bsnm{Ho}, \binits{A.}},
\bauthor{\bsnm{Hoffmann}, \binits{M.}},
\bauthor{\bsnm{Huang}, \binits{T.}},
\bauthor{\bsnm{Humble}, \binits{T.S.}},
\bauthor{\bsnm{Isakov}, \binits{S.V.}},
\bauthor{\bsnm{Jeffrey}, \binits{E.}},
\bauthor{\bsnm{Jiang}, \binits{Z.}},
\bauthor{\bsnm{Kafri}, \binits{D.}},
\bauthor{\bsnm{Kechedzhi}, \binits{K.}},
\bauthor{\bsnm{Kelly}, \binits{J.}},
\bauthor{\bsnm{Klimov}, \binits{P.V.}},
\bauthor{\bsnm{Knysh}, \binits{S.}},
\bauthor{\bsnm{Korotkov}, \binits{A.}},
\bauthor{\bsnm{Kostritsa}, \binits{F.}},
\bauthor{\bsnm{Landhuis}, \binits{D.}},
\bauthor{\bsnm{Lindmark}, \binits{M.}},
\bauthor{\bsnm{Lucero}, \binits{E.}},
\bauthor{\bsnm{Lyakh}, \binits{D.}},
\bauthor{\bsnm{Mandr{\`a}}, \binits{S.}},
\bauthor{\bsnm{McClean}, \binits{J.R.}},
\bauthor{\bsnm{McEwen}, \binits{M.}},
\bauthor{\bsnm{Megrant}, \binits{A.}},
\bauthor{\bsnm{Mi}, \binits{X.}},
\bauthor{\bsnm{Michielsen}, \binits{K.}},
\bauthor{\bsnm{Mohseni}, \binits{M.}},
\bauthor{\bsnm{Mutus}, \binits{J.}},
\bauthor{\bsnm{Naaman}, \binits{O.}},
\bauthor{\bsnm{Neeley}, \binits{M.}},
\bauthor{\bsnm{Neill}, \binits{C.}},
\bauthor{\bsnm{Niu}, \binits{M.Y.}},
\bauthor{\bsnm{Ostby}, \binits{E.}},
\bauthor{\bsnm{Petukhov}, \binits{A.}},
\bauthor{\bsnm{Platt}, \binits{J.C.}},
\bauthor{\bsnm{Quintana}, \binits{C.}},
\bauthor{\bsnm{Rieffel}, \binits{E.G.}},
\bauthor{\bsnm{Roushan}, \binits{P.}},
\bauthor{\bsnm{Rubin}, \binits{N.C.}},
\bauthor{\bsnm{Sank}, \binits{D.}},
\bauthor{\bsnm{Satzinger}, \binits{K.J.}},
\bauthor{\bsnm{Smelyanskiy}, \binits{V.}},
\bauthor{\bsnm{Sung}, \binits{K.J.}},
\bauthor{\bsnm{Trevithick}, \binits{M.D.}},
\bauthor{\bsnm{Vainsencher}, \binits{A.}},
\bauthor{\bsnm{Villalonga}, \binits{B.}},
\bauthor{\bsnm{White}, \binits{T.}},
\bauthor{\bsnm{Yao}, \binits{Z.J.}},
\bauthor{\bsnm{Yeh}, \binits{P.}},
\bauthor{\bsnm{Zalcman}, \binits{A.}},
\bauthor{\bsnm{Neven}, \binits{H.}},
\bauthor{\bsnm{Martinis}, \binits{J.M.}}:
\batitle{Quantum supremacy using a programmable superconducting processor}.
\bjtitle{Nature}
\bvolume{574}(\bissue{7779}),
\bfpage{505}--\blpage{510}
(\byear{2019})
\doiurl{10.1038/s41586-019-1666-5}
\end{barticle}
\endbibitem

\bibitem[\protect\citeauthoryear{Kim et~al.}{2023}]{Kim2023}
\begin{barticle}
\bauthor{\bsnm{Kim}, \binits{Y.}},
\bauthor{\bsnm{Eddins}, \binits{A.}},
\bauthor{\bsnm{Anand}, \binits{S.}},
\bauthor{\bsnm{Wei}, \binits{K.X.}},
\bauthor{\bsnm{Berg}, \binits{E.}},
\bauthor{\bsnm{Rosenblatt}, \binits{S.}},
\bauthor{\bsnm{Nayfeh}, \binits{H.}},
\bauthor{\bsnm{Wu}, \binits{Y.}},
\bauthor{\bsnm{Zaletel}, \binits{M.}},
\bauthor{\bsnm{Temme}, \binits{K.}},
\bauthor{\bsnm{Kandala}, \binits{A.}}:
\batitle{Evidence for the utility of quantum computing before fault tolerance}.
\bjtitle{Nature}
\bvolume{618}(\bissue{7965}),
\bfpage{500}--\blpage{505}
(\byear{2023})
\doiurl{10.1038/s41586-023-06096-3}
\end{barticle}
\endbibitem

\bibitem[\protect\citeauthoryear{Aboy et~al.}{2022}]{Aboy2022}
\begin{barticle}
\bauthor{\bsnm{Aboy}, \binits{M.}},
\bauthor{\bsnm{Minssen}, \binits{T.}},
\bauthor{\bsnm{Kop}, \binits{M.}}:
\batitle{Mapping the patent landscape of quantum technologies: Patenting trends, innovation and policy implications}.
\bjtitle{IIC - International Review of Intellectual Property and Competition Law}
\bvolume{53}(\bissue{6}),
\bfpage{853}--\blpage{882}
(\byear{2022})
\doiurl{10.1007/s40319-022-01209-3}
\end{barticle}
\endbibitem

\bibitem[\protect\citeauthoryear{{van Deventer, Oskar} et~al.}{2022}]{vanDeventer2022}
\begin{barticle}
\bauthor{\bsnm{{van Deventer, Oskar}}},
\bauthor{\bsnm{{Spethmann, Nicolas}}},
\bauthor{\bsnm{{Loeffler, Marius}}},
\bauthor{\bsnm{{Amoretti, Michele}}},
\bauthor{\bsnm{{van den Brink, Rob}}},
\bauthor{\bsnm{{Bruno, Natalia}}},
\bauthor{\bsnm{{Comi, Paolo}}},
\bauthor{\bsnm{{Farrugia, Noel}}},
\bauthor{\bsnm{{Gramegna, Marco}}},
\bauthor{\bsnm{{Jenet, Andreas}}},
\bauthor{\bsnm{{Kassenberg, Ben}}},
\bauthor{\bsnm{{Kozlowski, Wojciech}}},
\bauthor{\bsnm{{Länger, Thomas}}},
\bauthor{\bsnm{{Lindstrom, Tobias}}},
\bauthor{\bsnm{{Martin, Vicente}}},
\bauthor{\bsnm{{Neumann, Niels}}},
\bauthor{\bsnm{{Papadopoulos, Homer}}},
\bauthor{\bsnm{{Pascazio, Saverio}}},
\bauthor{\bsnm{{Peev, Momtchil}}},
\bauthor{\bsnm{{Pitwon, Richard}}},
\bauthor{\bsnm{{Rol, M. Adriaan}}},
\bauthor{\bsnm{{Traina, Paolo}}},
\bauthor{\bsnm{{Venderbosch, Pim}}},
\bauthor{\bsnm{{Wilhelm-Mauch, Frank K.}}}:
\batitle{Towards european standards for quantum technologies}.
\bjtitle{EPJ Quantum Technol.}
\bvolume{9}(\bissue{1}),
\bfpage{33}
(\byear{2022})
\doiurl{10.1140/epjqt/s40507-022-00150-1}
\end{barticle}
\endbibitem

\bibitem[\protect\citeauthoryear{MAZUR}{}]{mazurquantum}
\begin{botherref}
\oauthor{\bsnm{MAZUR}, \binits{T.}}:
Quantum technologies and standardization.
STRATEGIC PERSPECTIVES ON CYBERSECURITY MANAGEMENT AND PUBLIC POLICIES,
34
\end{botherref}
\endbibitem

\bibitem[\protect\citeauthoryear{Jenet}{2020}]{jenet2020standards4quantum}
\begin{bchapter}
\bauthor{\bsnm{Jenet}, \binits{A.}}:
\bctitle{Standards4quantum: making quantum technology ready for industry. putting science into standards}.
In: \bbtitle{Putting Science Into Standards (May 1, 2020). JRC Conference and Workshop Report}
(\byear{2020})
\end{bchapter}
\endbibitem

\bibitem[\protect\citeauthoryear{Sood and Pooja}{2024}]{Sood2024}
\begin{barticle}
\bauthor{\bsnm{Sood}, \binits{S.K.}},
\bauthor{\bsnm{Pooja}}:
\batitle{Quantum computing review: A decade of research}.
\bjtitle{IEEE Transactions on Engineering Management}
\bvolume{71},
\bfpage{6662}--\blpage{6676}
(\byear{2024})
\doiurl{10.1109/TEM.2023.3284689}
\end{barticle}
\endbibitem

\bibitem[\protect\citeauthoryear{Coccia and Roshani}{2024}]{coccia2024new}
\begin{barticle}
\bauthor{\bsnm{Coccia}, \binits{M.}},
\bauthor{\bsnm{Roshani}, \binits{S.}}:
\batitle{New directions in quantum computing technology}.
\bjtitle{Journal of the Knowledge Economy SSRN}
(\byear{2024})
\doiurl{10.2139/ssrn.4871417}
\end{barticle}
\endbibitem

\bibitem[\protect\citeauthoryear{Houssein et~al.}{2022}]{Houssein2022}
\begin{barticle}
\bauthor{\bsnm{Houssein}, \binits{E.H.}},
\bauthor{\bsnm{Abohashima}, \binits{Z.}},
\bauthor{\bsnm{Elhoseny}, \binits{M.}},
\bauthor{\bsnm{Mohamed}, \binits{W.M.}}:
\batitle{Machine learning in the quantum realm: The state-of-the-art, challenges, and future vision}.
\bjtitle{Expert Systems with Applications}
\bvolume{194},
\bfpage{116512}
(\byear{2022})
\doiurl{10.1016/j.eswa.2022.116512}
\end{barticle}
\endbibitem

\bibitem[\protect\citeauthoryear{Preskill}{2018}]{Preskill2018quantumcomputingin}
\begin{barticle}
\bauthor{\bsnm{Preskill}, \binits{J.}}:
\batitle{Quantum {C}omputing in the {NISQ} era and beyond}.
\bjtitle{{Quantum}}
\bvolume{2},
\bfpage{79}
(\byear{2018})
\doiurl{10.22331/q-2018-08-06-79}
\end{barticle}
\endbibitem

\bibitem[\protect\citeauthoryear{Bova et~al.}{2021}]{bova2021commercial}
\begin{barticle}
\bauthor{\bsnm{Bova}, \binits{F.}},
\bauthor{\bsnm{Goldfarb}, \binits{A.}},
\bauthor{\bsnm{Melko}, \binits{R.G.}}:
\batitle{Commercial applications of quantum computing}.
\bjtitle{EPJ quantum technology}
\bvolume{8}(\bissue{1}),
\bfpage{2}
(\byear{2021})
\end{barticle}
\endbibitem

\bibitem[\protect\citeauthoryear{Bayerstadler et~al.}{2021}]{bayerstadler2021industry}
\begin{barticle}
\bauthor{\bsnm{Bayerstadler}, \binits{A.}},
\bauthor{\bsnm{Becquin}, \binits{G.}},
\bauthor{\bsnm{Binder}, \binits{J.}},
\bauthor{\bsnm{Botter}, \binits{T.}},
\bauthor{\bsnm{Ehm}, \binits{H.}},
\bauthor{\bsnm{Ehmer}, \binits{T.}},
\bauthor{\bsnm{Erdmann}, \binits{M.}},
\bauthor{\bsnm{Gaus}, \binits{N.}},
\bauthor{\bsnm{Harbach}, \binits{P.}},
\bauthor{\bsnm{Hess}, \binits{M.}}, \betal:
\batitle{Industry quantum computing applications}.
\bjtitle{EPJ Quantum Technology}
\bvolume{8}(\bissue{1}),
\bfpage{25}
(\byear{2021})
\end{barticle}
\endbibitem

\bibitem[\protect\citeauthoryear{Greinert and Müller}{2024}]{greinert_2024_10976836}
\begin{botherref}
\oauthor{\bsnm{Greinert}, \binits{F.}},
\oauthor{\bsnm{Müller}, \binits{R.}}:
{European Competence Framework for Quantum Technologies}.
Zenodo
(2024).
\doiurl{10.5281/zenodo.10976836} .
\url{https://doi.org/10.5281/zenodo.10976836}
\end{botherref}
\endbibitem

\bibitem[\protect\citeauthoryear{S{\'a}ez-Ortu{\~n}o et~al.}{2024}]{saez2024quantum}
\begin{barticle}
\bauthor{\bsnm{S{\'a}ez-Ortu{\~n}o}, \binits{L.}},
\bauthor{\bsnm{Huertas-Garcia}, \binits{R.}},
\bauthor{\bsnm{Forgas-Coll}, \binits{S.}},
\bauthor{\bsnm{S{\'a}nchez-Garc{\'\i}a}, \binits{J.}},
\bauthor{\bsnm{Puertas-Prats}, \binits{E.}}:
\batitle{Quantum computing for market research}.
\bjtitle{Journal of Innovation \& Knowledge}
\bvolume{9}(\bissue{3}),
\bfpage{100510}
(\byear{2024})
\end{barticle}
\endbibitem

\bibitem[\protect\citeauthoryear{Carberry et~al.}{2021}]{CARBERRY20212065}
\begin{bchapter}
\bauthor{\bsnm{Carberry}, \binits{D.}},
\bauthor{\bsnm{Nourbakhsh}, \binits{A.}},
\bauthor{\bsnm{Karon}, \binits{J.}},
\bauthor{\bsnm{Jones}, \binits{M.N.}},
\bauthor{\bsnm{Jadidi}, \binits{M.}},
\bauthor{\bsnm{Shahriari}, \binits{K.}},
\bauthor{\bsnm{Beenfeldt}, \binits{C.}},
\bauthor{\bsnm{Andersson}, \binits{M.P.}},
\bauthor{\bsnm{Mansouri}, \binits{S.S.}}:
\bctitle{Building knowledge capacity for quantum computing in engineering education}.
In: \beditor{\bsnm{Türkay}, \binits{M.}},
\beditor{\bsnm{Gani}, \binits{R.}} (eds.)
\bbtitle{31st European Symposium on Computer Aided Process Engineering}.
\bsertitle{Computer Aided Chemical Engineering},
vol. \bseriesno{50},
pp. \bfpage{2065}--\blpage{2070}.
\bpublisher{Elsevier},
\blocation{Amsterdam, The Netherlands}
(\byear{2021}).
\doiurl{10.1016/B978-0-323-88506-5.50319-3} .
\burl{https://www.sciencedirect.com/science/article/pii/B9780323885065503193}
\end{bchapter}
\endbibitem

\bibitem[\protect\citeauthoryear{Seegerer et~al.}{2021}]{10.1145/3481312.3481348}
\begin{bchapter}
\bauthor{\bsnm{Seegerer}, \binits{S.}},
\bauthor{\bsnm{Michaeli}, \binits{T.}},
\bauthor{\bsnm{Romeike}, \binits{R.}}:
\bctitle{Quantum computing as a topic in computer science education}.
In: \bbtitle{Proceedings of the 16th Workshop in Primary and Secondary Computing Education}.
\bsertitle{WiPSCE '21}.
\bpublisher{Association for Computing Machinery},
\blocation{New York, NY, USA}
(\byear{2021}).
\doiurl{10.1145/3481312.3481348} .
\burl{https://doi.org/10.1145/3481312.3481348}
\end{bchapter}
\endbibitem

\bibitem[\protect\citeauthoryear{Goorney et~al.}{2023}]{goorney2023frameworkcurriculumtransformationquantum}
\begin{botherref}
\oauthor{\bsnm{Goorney}, \binits{S.}},
\oauthor{\bsnm{Bley}, \binits{J.}},
\oauthor{\bsnm{Heusler}, \binits{S.}},
\oauthor{\bsnm{Sherson}, \binits{J.}}:
A Framework for Curriculum Transformation in Quantum Information Science and Technology Education
(2023).
\url{https://arxiv.org/abs/2308.10371}
\end{botherref}
\endbibitem

\bibitem[\protect\citeauthoryear{Goorney et~al.}{2024}]{Goorney2024}
\begin{barticle}
\bauthor{\bsnm{Goorney}, \binits{S.}},
\bauthor{\bsnm{Sarantinou}, \binits{M.}},
\bauthor{\bsnm{Sherson}, \binits{J.}}:
\batitle{The quantum technology open master: widening access to the quantum industry}.
\bjtitle{EPJ Quantum Technology}
\bvolume{11}(\bissue{1}),
\bfpage{7}
(\byear{2024})
\doiurl{10.1140/epjqt/s40507-024-00217-1}
\end{barticle}
\endbibitem

\bibitem[\protect\citeauthoryear{Meyer et~al.}{2024}]{Meyer2024}
\begin{barticle}
\bauthor{\bsnm{Meyer}, \binits{J.C.}},
\bauthor{\bsnm{Passante}, \binits{G.}},
\bauthor{\bsnm{Pollock}, \binits{S.J.}},
\bauthor{\bsnm{Wilcox}, \binits{B.R.}}:
\batitle{Introductory quantum information science coursework at us institutions: content coverage}.
\bjtitle{EPJ Quantum Technology}
\bvolume{11}(\bissue{1}),
\bfpage{16}
(\byear{2024})
\doiurl{10.1140/epjqt/s40507-024-00226-0}
\end{barticle}
\endbibitem

\bibitem[\protect\citeauthoryear{Angara et~al.}{2020}]{angara2020quantum}
\begin{bchapter}
\bauthor{\bsnm{Angara}, \binits{P.P.}},
\bauthor{\bsnm{Stege}, \binits{U.}},
\bauthor{\bsnm{MacLean}, \binits{A.}}:
\bctitle{Quantum computing for high-school students an experience report}.
In: \bbtitle{2020 IEEE International Conference on Quantum Computing and Engineering (QCE)},
pp. \bfpage{323}--\blpage{329}
(\byear{2020}).
\bcomment{IEEE}
\end{bchapter}
\endbibitem

\bibitem[\protect\citeauthoryear{Angara et~al.}{2021}]{angara2021teaching}
\begin{barticle}
\bauthor{\bsnm{Angara}, \binits{P.P.}},
\bauthor{\bsnm{Stege}, \binits{U.}},
\bauthor{\bsnm{MacLean}, \binits{A.}},
\bauthor{\bsnm{M{\"u}ller}, \binits{H.A.}},
\bauthor{\bsnm{Markham}, \binits{T.}}:
\batitle{Teaching quantum computing to high-school-aged youth: A hands-on approach}.
\bjtitle{IEEE Transactions on Quantum Engineering}
\bvolume{3},
\bfpage{1}--\blpage{15}
(\byear{2021})
\end{barticle}
\endbibitem

\bibitem[\protect\citeauthoryear{Walsh et~al.}{2021}]{walsh2021piloting}
\begin{barticle}
\bauthor{\bsnm{Walsh}, \binits{J.A.}},
\bauthor{\bsnm{Fenech}, \binits{M.}},
\bauthor{\bsnm{Tucker}, \binits{D.L.}},
\bauthor{\bsnm{Riegle-Crumb}, \binits{C.}},
\bauthor{\bsnm{La~Cour}, \binits{B.R.}}:
\batitle{Piloting a full-year, optics-based high school course on quantum computing}.
\bjtitle{Physics education}
\bvolume{57}(\bissue{2}),
\bfpage{025010}
(\byear{2021})
\end{barticle}
\endbibitem

\bibitem[\protect\citeauthoryear{Merzel et~al.}{2024}]{Merzel2024}
\begin{barticle}
\bauthor{\bsnm{Merzel}, \binits{A.}},
\bauthor{\bsnm{Bitzenbauer}, \binits{P.}},
\bauthor{\bsnm{Krijtenburg-Lewerissa}, \binits{K.}},
\bauthor{\bsnm{Stadermann}, \binits{K.}},
\bauthor{\bsnm{Andreotti}, \binits{E.}},
\bauthor{\bsnm{Anttila}, \binits{D.}},
\bauthor{\bsnm{Bondani}, \binits{M.}},
\bauthor{\bsnm{Chiofalo}, \binits{M.L.M.}},
\bauthor{\bsnm{Faleti{\v{c}}}, \binits{S.}},
\bauthor{\bsnm{Frans}, \binits{R.}},
\bauthor{\bsnm{Goorney}, \binits{S.}},
\bauthor{\bsnm{Greinert}, \binits{F.}},
\bauthor{\bsnm{Jur{\v{c}}i{\'{c}}}, \binits{L.}},
\bauthor{\bsnm{Koupilov{\'a}}, \binits{Z.}},
\bauthor{\bsnm{Malgieri}, \binits{M.}},
\bauthor{\bsnm{M{\"u}ller}, \binits{R.}},
\bauthor{\bsnm{Onorato}, \binits{P.}},
\bauthor{\bsnm{Pospiech}, \binits{G.}},
\bauthor{\bsnm{Ubben}, \binits{M.}},
\bauthor{\bsnm{Woitzik}, \binits{A.}},
\bauthor{\bsnm{Pol}, \binits{H.}}:
\batitle{The core of secondary level quantum education: a multi-stakeholder perspective}.
\bjtitle{EPJ Quantum Technology}
\bvolume{11}(\bissue{1}),
\bfpage{27}
(\byear{2024})
\doiurl{10.1140/epjqt/s40507-024-00237-x}
\end{barticle}
\endbibitem

\bibitem[\protect\citeauthoryear{Langston}{}]{Microsoft2022}
\begin{botherref}
\oauthor{\bsnm{Langston}, \binits{J.}}:
In a historic milestone, Azure Quantum demonstrates formerly elusive physics needed to build scalable topological qubits.
\url{https://news.microsoft.com/source/features/innovation/azure-quantum-majorana-topological-qubit/}.
Accessed: 22.07.2024
\end{botherref}
\endbibitem

\bibitem[\protect\citeauthoryear{}{}]{Defianceetf2023}
\begin{botherref}
Quantum Supremacy: The Race Among Google, Microsoft, Intel, and IBM.
\url{https://www.defianceetfs.com/quantum-supremacy-the-race-among-google-microsoft-intel-and-ibm/#_ftn3}.
Accessed: 22.07.2024
\end{botherref}
\endbibitem

\bibitem[\protect\citeauthoryear{Kasneci et~al.}{2023}]{kasneci2023chatgpt}
\begin{barticle}
\bauthor{\bsnm{Kasneci}, \binits{E.}},
\bauthor{\bsnm{Se{\ss}ler}, \binits{K.}},
\bauthor{\bsnm{K{\"u}chemann}, \binits{S.}},
\bauthor{\bsnm{Bannert}, \binits{M.}},
\bauthor{\bsnm{Dementieva}, \binits{D.}},
\bauthor{\bsnm{Fischer}, \binits{F.}},
\bauthor{\bsnm{Gasser}, \binits{U.}},
\bauthor{\bsnm{Groh}, \binits{G.}},
\bauthor{\bsnm{G{\"u}nnemann}, \binits{S.}},
\bauthor{\bsnm{H{\"u}llermeier}, \binits{E.}}, \betal:
\batitle{Chatgpt for good? on opportunities and challenges of large language models for education}.
\bjtitle{Learning and individual differences}
\bvolume{103},
\bfpage{102274}
(\byear{2023})
\end{barticle}
\endbibitem

\bibitem[\protect\citeauthoryear{Ahmad et~al.}{2023}]{ahmad2023gen}
\begin{barticle}
\bauthor{\bsnm{Ahmad}, \binits{N.}},
\bauthor{\bsnm{Murugesan}, \binits{S.}},
\bauthor{\bsnm{Kshetri}, \binits{N.}}:
\batitle{Generative artificial intelligence and the education sector}.
\bjtitle{Computer}
\bvolume{56}(\bissue{6}),
\bfpage{72}--\blpage{76}
(\byear{2023})
\doiurl{10.1109/MC.2023.3263576}
\end{barticle}
\endbibitem

\bibitem[\protect\citeauthoryear{Krupp et~al.}{2024}]{krupp2024unreflected}
\begin{bchapter}
\bauthor{\bsnm{Krupp}, \binits{L.}},
\bauthor{\bsnm{Steinert}, \binits{S.}},
\bauthor{\bsnm{Kiefer-Emmanouilidis}, \binits{M.}},
\bauthor{\bsnm{Avila}, \binits{K.E.}},
\bauthor{\bsnm{Lukowicz}, \binits{P.}},
\bauthor{\bsnm{Kuhn}, \binits{J.}},
\bauthor{\bsnm{K{\"u}chemann}, \binits{S.}},
\bauthor{\bsnm{Karolus}, \binits{J.}}:
\bctitle{Unreflected acceptance--investigating the negative consequences of chatgpt-assisted problem solving in physics education}.
In: \bbtitle{HHAI 2024: Hybrid Human AI Systems for the Social Good},
pp. \bfpage{199}--\blpage{212}.
\bpublisher{IOS Press},
\blocation{Amsterdam, The Netherlands}
(\byear{2024})
\end{bchapter}
\endbibitem

\bibitem[\protect\citeauthoryear{Gregorcic and Pendrill}{2023}]{gregorcic2023chatgpt}
\begin{barticle}
\bauthor{\bsnm{Gregorcic}, \binits{B.}},
\bauthor{\bsnm{Pendrill}, \binits{A.-M.}}:
\batitle{Chatgpt and the frustrated socrates}.
\bjtitle{Physics Education}
\bvolume{58}(\bissue{3}),
\bfpage{035021}
(\byear{2023})
\end{barticle}
\endbibitem

\bibitem[\protect\citeauthoryear{Santos}{2023}]{santos2023enhancing}
\begin{botherref}
\oauthor{\bsnm{Santos}, \binits{R.P.d.}}:
Enhancing physics learning with chatgpt, bing chat, and bard as agents-to-think-with: A comparative case study.
arXiv preprint arXiv:2306.00724
(2023)
\end{botherref}
\endbibitem

\bibitem[\protect\citeauthoryear{Krupp et~al.}{2023}]{krupp2023challenges}
\begin{botherref}
\oauthor{\bsnm{Krupp}, \binits{L.}},
\oauthor{\bsnm{Steinert}, \binits{S.}},
\oauthor{\bsnm{Kiefer-Emmanouilidis}, \binits{M.}},
\oauthor{\bsnm{Avila}, \binits{K.E.}},
\oauthor{\bsnm{Lukowicz}, \binits{P.}},
\oauthor{\bsnm{Kuhn}, \binits{J.}},
\oauthor{\bsnm{K{\"u}chemann}, \binits{S.}},
\oauthor{\bsnm{Karolus}, \binits{J.}}:
Challenges and opportunities of moderating usage of large language models in education.
arXiv preprint arXiv:2312.14969
(2023)
\end{botherref}
\endbibitem

\bibitem[\protect\citeauthoryear{Sallam et~al.}{2023}]{sallam2023chatgpt}
\begin{botherref}
\oauthor{\bsnm{Sallam}, \binits{M.}},
\oauthor{\bsnm{Salim}, \binits{N.A.}},
\oauthor{\bsnm{Barakat}, \binits{M.}},
\oauthor{\bsnm{Ala'a}, \binits{B.}}:
Chatgpt applications in medical, dental, pharmacy, and public health education: A descriptive study highlighting the advantages and limitations.
Narra J
\textbf{3}(1)
(2023)
\end{botherref}
\endbibitem

\bibitem[\protect\citeauthoryear{Gregorcic and Polverini}{2024}]{gregorcic2024chatgpt}
\begin{botherref}
\oauthor{\bsnm{Gregorcic}, \binits{B.}},
\oauthor{\bsnm{Polverini}, \binits{G.}}:
Chatgpt-4 and the satisfied socrates.
arXiv preprint arXiv:2401.11987
(2024)
\end{botherref}
\endbibitem

\bibitem[\protect\citeauthoryear{{OpenAI}}{2022}]{openaichatgpt2022}
\begin{botherref}
\oauthor{\bsnm{{OpenAI}}}:
Introducing {ChatGPT} \url{https://openai.com/blog/chatgpt}
(2022)
\end{botherref}
\endbibitem

\bibitem[\protect\citeauthoryear{Latif et~al.}{2024}]{latif2024physicsassistant}
\begin{botherref}
\oauthor{\bsnm{Latif}, \binits{E.}},
\oauthor{\bsnm{Parasuraman}, \binits{R.}},
\oauthor{\bsnm{Zhai}, \binits{X.}}:
PhysicsAssistant: An LLM-Powered Interactive Learning Robot for Physics Lab Investigations
(2024)
\end{botherref}
\endbibitem

\bibitem[\protect\citeauthoryear{Wermelinger}{2023}]{wermelinger2023using}
\begin{bchapter}
\bauthor{\bsnm{Wermelinger}, \binits{M.}}:
\bctitle{Using github copilot to solve simple programming problems}.
In: \bbtitle{Proceedings of the 54th ACM Technical Symposium on Computer Science Education V. 1},
pp. \bfpage{172}--\blpage{178}
(\byear{2023})
\end{bchapter}
\endbibitem

\bibitem[\protect\citeauthoryear{G{\'o}mez-Rodr{\'\i}guez and Williams}{2023}]{gomez2023confederacy}
\begin{botherref}
\oauthor{\bsnm{G{\'o}mez-Rodr{\'\i}guez}, \binits{C.}},
\oauthor{\bsnm{Williams}, \binits{P.}}:
A confederacy of models: A comprehensive evaluation of llms on creative writing.
arXiv preprint arXiv:2310.08433
(2023)
\end{botherref}
\endbibitem

\bibitem[\protect\citeauthoryear{Thirunavukarasu et~al.}{2023}]{thirunavukarasu2023large}
\begin{barticle}
\bauthor{\bsnm{Thirunavukarasu}, \binits{A.J.}},
\bauthor{\bsnm{Ting}, \binits{D.S.J.}},
\bauthor{\bsnm{Elangovan}, \binits{K.}},
\bauthor{\bsnm{Gutierrez}, \binits{L.}},
\bauthor{\bsnm{Tan}, \binits{T.F.}},
\bauthor{\bsnm{Ting}, \binits{D.S.W.}}:
\batitle{Large language models in medicine}.
\bjtitle{Nature medicine}
\bvolume{29}(\bissue{8}),
\bfpage{1930}--\blpage{1940}
(\byear{2023})
\end{barticle}
\endbibitem

\bibitem[\protect\citeauthoryear{Team et~al.}{2024}]{geminiteam2024gemini}
\begin{botherref}
\oauthor{\bsnm{Team}, \binits{G.}},
\oauthor{\bsnm{Anil}, \binits{R.}},
\oauthor{\bsnm{Borgeaud}, \binits{S.}},
\oauthor{\bsnm{Alayrac}, \binits{J.-B.}},
\oauthor{\bsnm{Yu}, \binits{J.}},
\oauthor{\bsnm{al.}, \binits{R.S.}}:
Gemini: A Family of Highly Capable Multimodal Models
(2024)
\end{botherref}
\endbibitem

\bibitem[\protect\citeauthoryear{{Meta}}{2024}]{llama32024}
\begin{botherref}
\oauthor{\bsnm{{Meta}}}:
Introducing {Llama 3} \url{https://ai.meta.com/blog/meta-llama-3/}
(2024)
\end{botherref}
\endbibitem

\bibitem[\protect\citeauthoryear{{Microsoft}}{2023}]{microsoftcollab2023}
\begin{botherref}
\oauthor{\bsnm{{Microsoft}}}:
{Microsoft} and {OpenAI} extend partnership \url{https://blogs.microsoft.com/blog/2023/01/23/microsoftandopenaiextendpartnership/}
(2023)
\end{botherref}
\endbibitem

\bibitem[\protect\citeauthoryear{Parmar et~al.}{2024}]{parmar2024nemotron}
\begin{botherref}
\oauthor{\bsnm{Parmar}, \binits{J.}},
\oauthor{\bsnm{Prabhumoye}, \binits{S.}},
\oauthor{\bsnm{Jennings}, \binits{J.}},
\oauthor{\bsnm{Patwary}, \binits{M.}},
\oauthor{\bsnm{Subramanian}, \binits{S.}},
\oauthor{\bsnm{Su}, \binits{D.}},
\oauthor{\bsnm{Zhu}, \binits{C.}},
\oauthor{\bsnm{Narayanan}, \binits{D.}},
\oauthor{\bsnm{Jhunjhunwala}, \binits{A.}},
\oauthor{\bsnm{Dattagupta}, \binits{A.}}, et al.:
Nemotron-4 15b technical report.
arXiv preprint arXiv:2402.16819
(2024)
\end{botherref}
\endbibitem

\bibitem[\protect\citeauthoryear{Templeton et~al.}{2024}]{templeton2024scaling}
\begin{botherref}
\oauthor{\bsnm{Templeton}, \binits{A.}},
\oauthor{\bsnm{Conerly}, \binits{T.}},
\oauthor{\bsnm{Marcus}, \binits{J.}},
\oauthor{\bsnm{Lindsey}, \binits{J.}},
\oauthor{\bsnm{Bricken}, \binits{T.}},
\oauthor{\bsnm{Chen}, \binits{B.}},
\oauthor{\bsnm{Pearce}, \binits{A.}},
\oauthor{\bsnm{Citro}, \binits{C.}},
\oauthor{\bsnm{Ameisen}, \binits{E.}},
\oauthor{\bsnm{Jones}, \binits{A.}},
\oauthor{\bsnm{Cunningham}, \binits{H.}},
\oauthor{\bsnm{Turner}, \binits{N.L.}},
\oauthor{\bsnm{McDougall}, \binits{C.}},
\oauthor{\bsnm{MacDiarmid}, \binits{M.}},
\oauthor{\bsnm{Freeman}, \binits{C.D.}},
\oauthor{\bsnm{Sumers}, \binits{T.R.}},
\oauthor{\bsnm{Rees}, \binits{E.}},
\oauthor{\bsnm{Batson}, \binits{J.}},
\oauthor{\bsnm{Jermyn}, \binits{A.}},
\oauthor{\bsnm{Carter}, \binits{S.}},
\oauthor{\bsnm{Olah}, \binits{C.}},
\oauthor{\bsnm{Henighan}, \binits{T.}}:
Scaling monosemanticity: Extracting interpretable features from claude 3 sonnet.
Transformer Circuits Thread
(2024)
\end{botherref}
\endbibitem

\bibitem[\protect\citeauthoryear{OpenAI et~al.}{2024}]{openai2024gpt4}
\begin{botherref}
\oauthor{\bsnm{OpenAI}},
\oauthor{\bsnm{Achiam}, \binits{J.}},
\oauthor{\bsnm{Adler}, \binits{S.}},
\oauthor{\bsnm{Agarwal}, \binits{S.}},
\oauthor{\bsnm{Ahmad}, \binits{L.}},
\oauthor{\bsnm{al.}, \binits{I.A.}}:
GPT-4 Technical Report
(2024)
\end{botherref}
\endbibitem

\bibitem[\protect\citeauthoryear{Lewis et~al.}{2020}]{lewis2020retrieval}
\begin{barticle}
\bauthor{\bsnm{Lewis}, \binits{P.}},
\bauthor{\bsnm{Perez}, \binits{E.}},
\bauthor{\bsnm{Piktus}, \binits{A.}},
\bauthor{\bsnm{Petroni}, \binits{F.}},
\bauthor{\bsnm{Karpukhin}, \binits{V.}},
\bauthor{\bsnm{Goyal}, \binits{N.}},
\bauthor{\bsnm{K{\"u}ttler}, \binits{H.}},
\bauthor{\bsnm{Lewis}, \binits{M.}},
\bauthor{\bsnm{Yih}, \binits{W.-t.}},
\bauthor{\bsnm{Rockt{\"a}schel}, \binits{T.}}, \betal:
\batitle{Retrieval-augmented generation for knowledge-intensive nlp tasks}.
\bjtitle{Advances in Neural Information Processing Systems}
\bvolume{33},
\bfpage{9459}--\blpage{9474}
(\byear{2020})
\end{barticle}
\endbibitem

\bibitem[\protect\citeauthoryear{Gao et~al.}{2024}]{gao2024retrievalaugmented}
\begin{botherref}
\oauthor{\bsnm{Gao}, \binits{Y.}},
\oauthor{\bsnm{Xiong}, \binits{Y.}},
\oauthor{\bsnm{Gao}, \binits{X.}},
\oauthor{\bsnm{Jia}, \binits{K.}},
\oauthor{\bsnm{Pan}, \binits{J.}},
\oauthor{\bsnm{Bi}, \binits{Y.}},
\oauthor{\bsnm{Dai}, \binits{Y.}},
\oauthor{\bsnm{Sun}, \binits{J.}},
\oauthor{\bsnm{Wang}, \binits{M.}},
\oauthor{\bsnm{Wang}, \binits{H.}}:
Retrieval-Augmented Generation for Large Language Models: A Survey
(2024)
\end{botherref}
\endbibitem

\bibitem[\protect\citeauthoryear{Yan et~al.}{2024}]{https://doi.org/10.1111/bjet.13370}
\begin{barticle}
\bauthor{\bsnm{Yan}, \binits{L.}},
\bauthor{\bsnm{Sha}, \binits{L.}},
\bauthor{\bsnm{Zhao}, \binits{L.}},
\bauthor{\bsnm{Li}, \binits{Y.}},
\bauthor{\bsnm{Martinez-Maldonado}, \binits{R.}},
\bauthor{\bsnm{Chen}, \binits{G.}},
\bauthor{\bsnm{Li}, \binits{X.}},
\bauthor{\bsnm{Jin}, \binits{Y.}},
\bauthor{\bsnm{Gašević}, \binits{D.}}:
\batitle{Practical and ethical challenges of large language models in education: A systematic scoping review}.
\bjtitle{British Journal of Educational Technology}
\bvolume{55}(\bissue{1}),
\bfpage{90}--\blpage{112}
(\byear{2024})
\doiurl{10.1111/bjet.13370}
{\href{https://arxiv.org/abs/https://bera-journals.onlinelibrary.wiley.com/doi/pdf/10.1111/bjet.13370}{{https://bera-journals.onlinelibrary.wiley.com/doi/pdf/10.1111/bjet.13370}}}
\end{barticle}
\endbibitem

\bibitem[\protect\citeauthoryear{Belland}{2017}]{belland2017instructional}
\begin{bbook}
\bauthor{\bsnm{Belland}, \binits{B.R.}}:
\bbtitle{Instructional Scaffolding in STEM Education: Strategies and Efficacy Evidence}.
\bpublisher{Springer},
\blocation{London, United Kingdom}
(\byear{2017})
\end{bbook}
\endbibitem

\bibitem[\protect\citeauthoryear{Schartel}{2012}]{SCHARTEL201277}
\begin{barticle}
\bauthor{\bsnm{Schartel}, \binits{S.A.}}:
\batitle{Giving feedback – an integral part of education}.
\bjtitle{Best Practice \& Research Clinical Anaesthesiology}
\bvolume{26}(\bissue{1}),
\bfpage{77}--\blpage{87}
(\byear{2012})
\doiurl{10.1016/j.bpa.2012.02.003} .
\bcomment{Challenges in Anaesthesia Education}
\end{barticle}
\endbibitem

\bibitem[\protect\citeauthoryear{Bajaj and Sharma}{2018}]{BAJAJ2018834}
\begin{barticle}
\bauthor{\bsnm{Bajaj}, \binits{R.}},
\bauthor{\bsnm{Sharma}, \binits{V.}}:
\batitle{Smart education with artificial intelligence based determination of learning styles}.
\bjtitle{Procedia Computer Science}
\bvolume{132},
\bfpage{834}--\blpage{842}
(\byear{2018})
\doiurl{10.1016/j.procs.2018.05.095} .
\bcomment{International Conference on Computational Intelligence and Data Science}
\end{barticle}
\endbibitem

\bibitem[\protect\citeauthoryear{Gan et~al.}{2023}]{10386291}
\begin{bchapter}
\bauthor{\bsnm{Gan}, \binits{W.}},
\bauthor{\bsnm{Qi}, \binits{Z.}},
\bauthor{\bsnm{Wu}, \binits{J.}},
\bauthor{\bsnm{Lin}, \binits{J.C.-W.}}:
\bctitle{Large language models in education: Vision and opportunities}.
In: \bbtitle{2023 IEEE International Conference on Big Data (BigData)},
pp. \bfpage{4776}--\blpage{4785}
(\byear{2023}).
\doiurl{10.1109/BigData59044.2023.10386291}
\end{bchapter}
\endbibitem

\bibitem[\protect\citeauthoryear{Zhu et~al.}{2016}]{Zhu2016}
\begin{barticle}
\bauthor{\bsnm{Zhu}, \binits{Z.-T.}},
\bauthor{\bsnm{Yu}, \binits{M.-H.}},
\bauthor{\bsnm{Riezebos}, \binits{P.}}:
\batitle{A research framework of smart education}.
\bjtitle{Smart Learning Environments}
\bvolume{3}(\bissue{1}),
\bfpage{4}
(\byear{2016})
\doiurl{10.1186/s40561-016-0026-2}
\end{barticle}
\endbibitem

\bibitem[\protect\citeauthoryear{Weidinger et~al.}{2022}]{10.1145/3531146.3533088}
\begin{bchapter}
\bauthor{\bsnm{Weidinger}, \binits{L.}},
\bauthor{\bsnm{Uesato}, \binits{J.}},
\bauthor{\bsnm{Rauh}, \binits{M.}},
\bauthor{\bsnm{Griffin}, \binits{C.}},
\bauthor{\bsnm{Huang}, \binits{P.-S.}},
\bauthor{\bsnm{Mellor}, \binits{J.}},
\bauthor{\bsnm{Glaese}, \binits{A.}},
\bauthor{\bsnm{Cheng}, \binits{M.}},
\bauthor{\bsnm{Balle}, \binits{B.}},
\bauthor{\bsnm{Kasirzadeh}, \binits{A.}},
\bauthor{\bsnm{Biles}, \binits{C.}},
\bauthor{\bsnm{Brown}, \binits{S.}},
\bauthor{\bsnm{Kenton}, \binits{Z.}},
\bauthor{\bsnm{Hawkins}, \binits{W.}},
\bauthor{\bsnm{Stepleton}, \binits{T.}},
\bauthor{\bsnm{Birhane}, \binits{A.}},
\bauthor{\bsnm{Hendricks}, \binits{L.A.}},
\bauthor{\bsnm{Rimell}, \binits{L.}},
\bauthor{\bsnm{Isaac}, \binits{W.}},
\bauthor{\bsnm{Haas}, \binits{J.}},
\bauthor{\bsnm{Legassick}, \binits{S.}},
\bauthor{\bsnm{Irving}, \binits{G.}},
\bauthor{\bsnm{Gabriel}, \binits{I.}}:
\bctitle{Taxonomy of risks posed by language models}.
In: \bbtitle{Proceedings of the 2022 ACM Conference on Fairness, Accountability, and Transparency}.
\bsertitle{FAccT '22},
pp. \bfpage{214}--\blpage{229}.
\bpublisher{Association for Computing Machinery},
\blocation{New York, NY, USA}
(\byear{2022}).
\doiurl{10.1145/3531146.3533088} .
\burl{https://doi.org/10.1145/3531146.3533088}
\end{bchapter}
\endbibitem

\bibitem[\protect\citeauthoryear{Milano et~al.}{2023}]{Milano2023}
\begin{barticle}
\bauthor{\bsnm{Milano}, \binits{S.}},
\bauthor{\bsnm{McGrane}, \binits{J.A.}},
\bauthor{\bsnm{Leonelli}, \binits{S.}}:
\batitle{Large language models challenge the future of higher education}.
\bjtitle{Nature Machine Intelligence}
\bvolume{5}(\bissue{4}),
\bfpage{333}--\blpage{334}
(\byear{2023})
\doiurl{10.1038/s42256-023-00644-2}
\end{barticle}
\endbibitem

\bibitem[\protect\citeauthoryear{Kosch et~al.}{2023}]{kosch2023placebo}
\begin{barticle}
\bauthor{\bsnm{Kosch}, \binits{T.}},
\bauthor{\bsnm{Welsch}, \binits{R.}},
\bauthor{\bsnm{Chuang}, \binits{L.}},
\bauthor{\bsnm{Schmidt}, \binits{A.}}:
\batitle{The placebo effect of artificial intelligence in human--computer interaction}.
\bjtitle{ACM Transactions on Computer-Human Interaction}
\bvolume{29}(\bissue{6}),
\bfpage{1}--\blpage{32}
(\byear{2023})
\end{barticle}
\endbibitem

\bibitem[\protect\citeauthoryear{Kloft et~al.}{2024}]{kloft2024ai}
\begin{bchapter}
\bauthor{\bsnm{Kloft}, \binits{A.M.}},
\bauthor{\bsnm{Welsch}, \binits{R.}},
\bauthor{\bsnm{Kosch}, \binits{T.}},
\bauthor{\bsnm{Villa}, \binits{S.}}:
\bctitle{{"AI enhances our performance, I have no doubt this one will do the same": The Placebo Effect Is Robust to Negative Descriptions of AI}}.
In: \bbtitle{{Proceedings of the 2024 CHI Conference on Human Factors in Computing Systems}}.
\bsertitle{CHI '24}.
\bpublisher{ACM},
\blocation{New York, NY, USA}
(\byear{2024}).
\doiurl{10.1145/3613904.3642633}
\end{bchapter}
\endbibitem

\bibitem[\protect\citeauthoryear{Hu et~al.}{2024}]{Hu_2024}
\begin{barticle}
\bauthor{\bsnm{Hu}, \binits{P.}},
\bauthor{\bsnm{Li}, \binits{Y.}},
\bauthor{\bsnm{Mong}, \binits{R.S.K.}},
\bauthor{\bsnm{Singh}, \binits{C.}}:
\batitle{Student understanding of the bloch sphere}.
\bjtitle{European Journal of Physics}
\bvolume{45}(\bissue{2}),
\bfpage{025705}
(\byear{2024})
\doiurl{10.1088/1361-6404/ad2393}
\end{barticle}
\endbibitem

\bibitem[\protect\citeauthoryear{Boynton et~al.}{2013}]{boynton2013exploring}
\begin{barticle}
\bauthor{\bsnm{Boynton}, \binits{M.H.}},
\bauthor{\bsnm{Portnoy}, \binits{D.B.}},
\bauthor{\bsnm{Johnson}, \binits{B.T.}}:
\batitle{Exploring the ethics and psychological impact of deception in psychological research}.
\bjtitle{IRB}
\bvolume{35}(\bissue{2}),
\bfpage{7}
(\byear{2013})
\end{barticle}
\endbibitem

\bibitem[\protect\citeauthoryear{Girden}{1992}]{girden1992anova}
\begin{bbook}
\bauthor{\bsnm{Girden}, \binits{E.R.}}:
\bbtitle{ANOVA: Repeated Measures}
vol. \bseriesno{84}.
\bpublisher{Sage},
\blocation{London, United Kingdom}
(\byear{1992})
\end{bbook}
\endbibitem

\bibitem[\protect\citeauthoryear{Wobbrock et~al.}{2011}]{wobbrock2011aligned}
\begin{bchapter}
\bauthor{\bsnm{Wobbrock}, \binits{J.O.}},
\bauthor{\bsnm{Findlater}, \binits{L.}},
\bauthor{\bsnm{Gergle}, \binits{D.}},
\bauthor{\bsnm{Higgins}, \binits{J.J.}}:
\bctitle{The aligned rank transform for nonparametric factorial analyses using only anova procedures}.
In: \bbtitle{Proceedings of the SIGCHI Conference on Human Factors in Computing Systems},
pp. \bfpage{143}--\blpage{146}
(\byear{2011})
\end{bchapter}
\endbibitem

\bibitem[\protect\citeauthoryear{Jung et~al.}{2023}]{jung2023chatgpt}
\begin{barticle}
\bauthor{\bsnm{Jung}, \binits{L.B.}},
\bauthor{\bsnm{Gudera}, \binits{J.A.}},
\bauthor{\bsnm{Wiegand}, \binits{T.L.}},
\bauthor{\bsnm{Allmendinger}, \binits{S.}},
\bauthor{\bsnm{Dimitriadis}, \binits{K.}},
\bauthor{\bsnm{Koerte}, \binits{I.K.}}:
\batitle{Chatgpt passes german state examination in medicine with picture questions omitted}.
\bjtitle{Deutsches {\"A}rzteblatt International}
\bvolume{120}(\bissue{21-22}),
\bfpage{373}
(\byear{2023})
\end{barticle}
\endbibitem

\bibitem[\protect\citeauthoryear{Lee et~al.}{2023}]{lee2023few}
\begin{botherref}
\oauthor{\bsnm{Lee}, \binits{U.}},
\oauthor{\bsnm{Jung}, \binits{H.}},
\oauthor{\bsnm{Jeon}, \binits{Y.}},
\oauthor{\bsnm{Sohn}, \binits{Y.}},
\oauthor{\bsnm{Hwang}, \binits{W.}},
\oauthor{\bsnm{Moon}, \binits{J.}},
\oauthor{\bsnm{Kim}, \binits{H.}}:
Few-shot is enough: exploring chatgpt prompt engineering method for automatic question generation in english education.
Education and Information Technologies,
1--33
(2023)
\end{botherref}
\endbibitem

\bibitem[\protect\citeauthoryear{Zheng et~al.}{2024}]{zheng2024judging}
\begin{botherref}
\oauthor{\bsnm{Zheng}, \binits{L.}},
\oauthor{\bsnm{Chiang}, \binits{W.-L.}},
\oauthor{\bsnm{Sheng}, \binits{Y.}},
\oauthor{\bsnm{Zhuang}, \binits{S.}},
\oauthor{\bsnm{Wu}, \binits{Z.}},
\oauthor{\bsnm{Zhuang}, \binits{Y.}},
\oauthor{\bsnm{Lin}, \binits{Z.}},
\oauthor{\bsnm{Li}, \binits{Z.}},
\oauthor{\bsnm{Li}, \binits{D.}},
\oauthor{\bsnm{Xing}, \binits{E.}}, et al.:
Judging llm-as-a-judge with mt-bench and chatbot arena.
Advances in Neural Information Processing Systems
\textbf{36}
(2024)
\end{botherref}
\endbibitem

\bibitem[\protect\citeauthoryear{Krathwohl}{2002}]{krathwohl2002revision}
\begin{barticle}
\bauthor{\bsnm{Krathwohl}, \binits{D.R.}}:
\batitle{A revision of bloom's taxonomy: An overview}.
\bjtitle{Theory into practice}
\bvolume{41}(\bissue{4}),
\bfpage{212}--\blpage{218}
(\byear{2002})
\end{barticle}
\endbibitem

\bibitem[\protect\citeauthoryear{Erickson}{2002}]{erickson2002concept}
\begin{bbook}
\bauthor{\bsnm{Erickson}, \binits{H.L.}}:
\bbtitle{Concept-based Curriculum and Instruction: Teaching Beyond the Facts}.
\bpublisher{Corwin Press},
\blocation{Thousand Oaks, CA}
(\byear{2002})
\end{bbook}
\endbibitem

\bibitem[\protect\citeauthoryear{Tonmoy et~al.}{2024}]{tonmoy2024comprehensive}
\begin{botherref}
\oauthor{\bsnm{Tonmoy}, \binits{S.}},
\oauthor{\bsnm{Zaman}, \binits{S.}},
\oauthor{\bsnm{Jain}, \binits{V.}},
\oauthor{\bsnm{Rani}, \binits{A.}},
\oauthor{\bsnm{Rawte}, \binits{V.}},
\oauthor{\bsnm{Chadha}, \binits{A.}},
\oauthor{\bsnm{Das}, \binits{A.}}:
A comprehensive survey of hallucination mitigation techniques in large language models.
arXiv preprint arXiv:2401.01313
(2024)
\end{botherref}
\endbibitem

\bibitem[\protect\citeauthoryear{Pithers and Soden}{2000}]{pithers2000critical}
\begin{barticle}
\bauthor{\bsnm{Pithers}, \binits{R.T.}},
\bauthor{\bsnm{Soden}, \binits{R.}}:
\batitle{Critical thinking in education: A review}.
\bjtitle{Educational research}
\bvolume{42}(\bissue{3}),
\bfpage{237}--\blpage{249}
(\byear{2000})
\end{barticle}
\endbibitem

\bibitem[\protect\citeauthoryear{Villa et~al.}{2023}]{villa2023placebo}
\begin{barticle}
\bauthor{\bsnm{Villa}, \binits{S.}},
\bauthor{\bsnm{Kosch}, \binits{T.}},
\bauthor{\bsnm{Grelka}, \binits{F.}},
\bauthor{\bsnm{Schmidt}, \binits{A.}},
\bauthor{\bsnm{Welsch}, \binits{R.}}:
\batitle{The placebo effect of human augmentation: Anticipating cognitive augmentation increases risk-taking behavior}.
\bjtitle{Computers in Human Behavior}
\bvolume{146},
\bfpage{107787}
(\byear{2023})
\end{barticle}
\endbibitem

\bibitem[\protect\citeauthoryear{Hong}{1999}]{HONG1999431}
\begin{barticle}
\bauthor{\bsnm{Hong}, \binits{E.}}:
\batitle{Test anxiety, perceived test difficulty, and test performance: temporal patterns of their effects}.
\bjtitle{Learning and Individual Differences}
\bvolume{11}(\bissue{4}),
\bfpage{431}--\blpage{447}
(\byear{1999})
\doiurl{10.1016/S1041-6080(99)80012-0}
\end{barticle}
\endbibitem

\bibitem[\protect\citeauthoryear{Kukla}{1974}]{KUKLA1974374}
\begin{barticle}
\bauthor{\bsnm{Kukla}, \binits{A.}}:
\batitle{Performance as a function of resultant achievement motivation (perceived ability) and perceived difficulty}.
\bjtitle{Journal of Research in Personality}
\bvolume{7}(\bissue{4}),
\bfpage{374}--\blpage{383}
(\byear{1974})
\doiurl{10.1016/0092-6566(74)90059-2}
\end{barticle}
\endbibitem

\bibitem[\protect\citeauthoryear{Sarrazin et~al.}{2002}]{sarrazin2002exerted}
\begin{barticle}
\bauthor{\bsnm{Sarrazin}, \binits{P.}},
\bauthor{\bsnm{Roberts}, \binits{G.}},
\bauthor{\bsnm{Cury}, \binits{F.}},
\bauthor{\bsnm{Biddle}, \binits{S.}},
\bauthor{\bsnm{Famose}, \binits{J.-P.}}:
\batitle{Exerted effort and performance in climbing among boys: The influence of achievement goals, perceived ability, and task difficulty}.
\bjtitle{Research Quarterly for Exercise and Sport}
\bvolume{73}(\bissue{4}),
\bfpage{425}--\blpage{436}
(\byear{2002})
\end{barticle}
\endbibitem

\bibitem[\protect\citeauthoryear{Christ et~al.}{2024}]{christ2024mathwell}
\begin{botherref}
\oauthor{\bsnm{Christ}, \binits{B.R.}},
\oauthor{\bsnm{Kropko}, \binits{J.}},
\oauthor{\bsnm{Hartvigsen}, \binits{T.}}:
Mathwell: Generating educational math word problems at scale.
arXiv preprint arXiv:2402.15861
(2024)
\end{botherref}
\endbibitem

\bibitem[\protect\citeauthoryear{Huang et~al.}{2024}]{huang2024survey}
\begin{barticle}
\bauthor{\bsnm{Huang}, \binits{X.}},
\bauthor{\bsnm{Ruan}, \binits{W.}},
\bauthor{\bsnm{Huang}, \binits{W.}},
\bauthor{\bsnm{Jin}, \binits{G.}},
\bauthor{\bsnm{Dong}, \binits{Y.}},
\bauthor{\bsnm{Wu}, \binits{C.}},
\bauthor{\bsnm{Bensalem}, \binits{S.}},
\bauthor{\bsnm{Mu}, \binits{R.}},
\bauthor{\bsnm{Qi}, \binits{Y.}},
\bauthor{\bsnm{Zhao}, \binits{X.}}, \betal:
\batitle{A survey of safety and trustworthiness of large language models through the lens of verification and validation}.
\bjtitle{Artificial Intelligence Review}
\bvolume{57}(\bissue{7}),
\bfpage{175}
(\byear{2024})
\end{barticle}
\endbibitem

\bibitem[\protect\citeauthoryear{Holmes et~al.}{2013}]{holmes2013data}
\begin{botherref}
\oauthor{\bsnm{Holmes}, \binits{V.-L.}},
\oauthor{\bsnm{Miedema}, \binits{C.}},
\oauthor{\bsnm{Nieuwkoop}, \binits{L.}},
\oauthor{\bsnm{Haugen}, \binits{N.}}:
Data-driven intervention: correcting mathematics students' misconceptions, not mistakes.
The Mathematics Educator
\textbf{23}(1)
(2013)
\end{botherref}
\endbibitem

\bibitem[\protect\citeauthoryear{Amirizaniani et~al.}{2024}]{amirizaniani2024developing}
\begin{botherref}
\oauthor{\bsnm{Amirizaniani}, \binits{M.}},
\oauthor{\bsnm{Yao}, \binits{J.}},
\oauthor{\bsnm{Lavergne}, \binits{A.}},
\oauthor{\bsnm{Okada}, \binits{E.S.}},
\oauthor{\bsnm{Chadha}, \binits{A.}},
\oauthor{\bsnm{Roosta}, \binits{T.}},
\oauthor{\bsnm{Shah}, \binits{C.}}:
Developing a framework for auditing large language models using human-in-the-loop.
arXiv preprint arXiv:2402.09346
(2024)
\end{botherref}
\endbibitem

\end{thebibliography}
